\documentclass[journal=jacsat,manuscript=article]{achemso}

\usepackage[version=3]{mhchem} 
\usepackage{subfigure}
\usepackage{verbatim}
\usepackage{multirow}
\usepackage{multicol}
\usepackage[breaklinks]{hyperref}
\usepackage{color}

\usepackage{xr}
\usepackage{xcite}
\makeatletter
\newcommand*{\addFileDependency}[1]{%
  \typeout{(#1)}%
  \@addtofilelist{#1}%
  \IfFileExists{#1}{}{\typeout{No file #1.}}%
}
\makeatother

\author{Roshini Jayabalan}%
\affiliation[University of Augsburg]{Institute of Physics, Universität Augsburg, 86135 Augsburg, Germany}
\author{Girish K. Hanumantharaju}%
\affiliation[University of Augsburg]{Institute of Physics, Universität Augsburg, 86135 Augsburg, Germany}
\author{Theresa Hettiger}%
\affiliation[University of Tübingen]{Institute for Physical und Theoretical Chemistry, Universität T\"{u}bingen, 72076 T\"{u}bingen, Germany}%
\author{Arup Sarkar}%
\affiliation[Max Planck Institute for Polymer Research]{Max Planck Institute for Polymer Research, 55128 Mainz, Germany}
\author{Fengshuo Zu}%
\affiliation[Institute of Physics and IRIS Adlershof]{Helmholtz-Zentrum Berlin for Materials und Energy GmbH, 12489 Berlin, Germany}
\author{Aladin Ullrich}%
\affiliation[University of Augsburg]{Institute of Physics, Universität Augsburg, 86135 Augsburg, Germany}
\author{Norbert Koch}%
\affiliation[Institute of Physics and IRIS Adlershof]{Helmholtz-Zentrum Berlin for Materials und Energy GmbH, 12489 Berlin, Germany}
\alsoaffiliation{Institute of Physics \& Center for the Science of Materials Berlin, Humboldt-Universität zu Berlin, 12489 Berlin, Germany}
\author{Denis Andrienko}%
\affiliation[Max Planck Institute for Polymer Research]{Max Planck Institute for Polymer Research, 55128 Mainz, Germany}%
\author{Marcus Scheele}%
\affiliation[University of Tübingen]{Institute for Physical und Theoretical Chemistry, Universität T\"{u}bingen, 72076 T\"{u}bingen, Germany}%
\author{Wolfgang Br\"utting}%
\email{wolfgang.bruetting@physik.uni-augsburg.de}%
\affiliation[University of Augsburg]{Institute of Physics, Universität Augsburg, 86135 Augsburg, Germany}%

\title[An \textsf{achemso} demo]
 {Optimizing carrier balance in CsPbBr\textsubscript{3} nanocrystal LEDs: The role of alkyl ligands and polar electron transport layers}

\abbreviations{LHP, NC, PLQY, ETL, PeLED, SOP, DDABr}
\SectionNumbersOn

\begin{document}





\begin{abstract}

The study of lead halide perovskite nanocrystal based light-emitting diodes (LEDs) has advanced significantly, with notable improvements in stability and optical properties. However, optimizing charge carrier injection and transport remains a challenge. Efficient electroluminescence requires a balanced transport of both holes and electrons within the emitting material.
Here, we investigate cubic CsPbBr\textsubscript{3} nanocrystals passivated with oleylamine and oleic acid, comparing them to ligand-exchanged nanocrystals with didodecyldimethylammonium bromide (DDABr). Nuclear magnetic resonance spectroscopy and transmission electron microscopy confirm successful ligand exchange, revealing reduced ligand coverage in DDABr-treated nanocrystals. Photoelectron spectroscopy, spectroelectrochemistry, and single-carrier devices indicate improved hole injection in DDABr-capped nanocrystals. Density functional theory calculations further reveal the influence of ligand type and coverage on energy levels, with oleic acid introducing localized states in native nanocrystals. Additionally, incorporation of a polar electron transport layer (ETL) enhances LED performance by over an order of magnitude in DDABr-capped nanocrystals, driven by improved charge balance arising from the spontaneous orientation polarization (SOP) of the ETL.  These findings highlight the critical role of ligand selection, passivation degree, and charge transport control by the adjacent organic transport layers in optimizing LED efficiency.

\end{abstract}

\section{Introduction}
Lead halide perovskite nanocrystals (LHP-NCs) have gained significant attention in the field of optoelectronics due to their extraordinary optical and electronic properties, which make them highly promising candidates for light-emitting applications, particularly in perovskite-based light-emitting diodes (PeLEDs)\cite{Protesescu2015}. The typical structure of these materials is denoted by the formula ABX\textsubscript{3}, where A is an organic or inorganic monovalent cation (e.g., methylammonium (MA\textsuperscript{+}), formamidinium (FA\textsuperscript{+}), or cesium (Cs\textsuperscript{+})), B is a divalent metal cation (typically lead, Pb\textsuperscript{2+}), and X is a halide anion (Cl\textsuperscript{-}, Br\textsuperscript{-}, or I\textsuperscript{-}). By varying the halide composition, the emission wavelength of LHP-NCs can be precisely tuned across the entire visible spectrum. This flexibility makes them attractive for full-color displays with narrow emission linewidths (full-width at half-maximum, (FWHM) of less than 20 nm), providing high color purity while exhibiting high photoluminescent quantum yield (PLQY) close to unity\cite{Protesescu2015}. 

One of the key advantages of LHP-NCs is their solution-based colloidal synthesis, which allows for low-cost, scalable manufacturing methods like spin-coating or inkjet printing\cite{Cohen2022,Lee2021,Ye2022}. This positions them as an alternative to other materials like organic emitters\cite{Shibata2015} and quantum dots (QDs)\cite{Walied2021}, which often require more complex or expensive deposition and synthesis techniques. Another attractive property of LHP-NCs is their defect tolerance\cite{Huang2017}. Their ability to maintain high PLQYs despite the presence of crystal defects reduces the need for extensive defect engineering, which is often required in other semiconductors like II-VI quantum dots or III-V materials\cite{Walied2021,Aladakov2019}. These properties make LHP-NCs highly attractive for next-generation displays, lighting, and lasers, as well as other optoelectronic applications. However, despite their outstanding properties, there are still several challenges that must be addressed to fully realize their application potential \cite{Fosse2022,Fuiza2023}. 

A significant bottleneck in leveraging LHP-NCs for device applications lies in their imbalanced charge transport characteristics. Efficient charge injection in LHP-NC-based devices is often hindered by the insulating nature of the long alkyl chain ligands typically oleylamine (OAm) and oleic acid (OA)  used during synthesis (Fig. \ref{Ligands}a \& b). These ligands not only act as capping agents and aid precursor dissolution for colloidal NC formation \cite{Otero2021}, but their structural features such as the bent conformation of oleic acid due to the cis-double bond at the 9th carbon reduce van der Waals interactions among hydrocarbon chains, enhancing NC dispersibility and influencing their morphology and size \cite{Fuiza2023,Trizio2023,Choi2024}. While these long-chain ligands ensure colloidal stability and support high PLQY, they simultaneously impede charge transport in PeLEDs. Additionally, the inherent imbalance between electron and hole transport can further reduce PeLED efficiency, likely due to enhanced Auger recombination losses \cite{Deschler2016}. One promising strategy to address these limitations involves ligand exchange, which directly tunes the optoelectronic properties of the NCs \cite{Fuiza2023}. This typically entails replacing insulating ligands with shorter or more conductive alternatives \cite{Choi2024}, or introducing multidentate and zwitterionic ligands to strengthen surface binding \cite{Krieg2018}. Although parameters like alkyl chain length and ligand coverage have been systematically investigated for their effects on PeLED performance \cite{Dai2024,Kumar2019}, the specific role of the ligand’s anchoring group remains comparatively underexplored. 
  
  Didodecyldimethylammonium bromide (DDABr)  (Fig.~\ref{Ligands}c) has emerged as one of the most promising candidates for ligand exchange in lead halide perovskite nanocrystals for LED applications \cite{Shynkarenko2019}. DDABr is a quaternary ammonium salt with a bromide anion (Br\textsuperscript{-}) and a large organic cation (DDA\textsuperscript{+}) that offers multiple advantages in surface passivation and device performance. For example, the Br\textsuperscript{-} anion can fill halide vacancies in the perovskite structure, reducing surface defects that would otherwise lead to non-radiative recombination and lower PLQY, while the double alkyl chains of DDA\textsuperscript{+} cations provide steric protection to the perovskite NCs\cite{Zaccaria2022,Fuiza2023}, preventing them from aggregating and maintaining colloidal stability. The length of the carbon chain is reduced in comparision to OA/OAm, implying a potentially closer packing of NCs with improved charge transport. In addition to this, several reports suggest an improved operational stability and higher efficiency in PeLEDs while employing DDABr ligands onto their NCs\cite{Pan2016,Shynkarenko2019,Sun2024,Uribe2024,Hung2022,Yang2021}. 
  
  In this work, we address the importance of the anchoring group in these alkyl ligands attaching to the NC surface and its effect on carrier injection and transport in PeLEDs. Specifically, we investigate changes in film morphology, electronic properties and carrier transport following ligand exchange on  \textquotedblleft native\textquotedblright,  i.e.,  CsPbBr\textsubscript{3} NCs originally capped with OA and OAm, and comparing them to DDABr-exchanged NCs. Quantification on the ligand coverage, the NC packing and electrical accessibility of the NCs is supported by transmission electron spectroscopy (TEM), nuclear magnetic resonance (NMR) spectroscopy and spectroelectrochemistry. The origin for the enhanced performance in PeLEDs with DDABr capped NCs is revealed by photoelectron spectroscopy and single carrier devices in combination with density functional studies (DFT) providing insights into the ligands' influence on the electronic properties of the NCs. In addition to ligand influence, the impact of spontaneous orientation polarization (SOP), particularly when using electron transport layers (ETLs) with a significant permanent dipole moment, further improves charge carrier balance in PeLEDs.


\section{Results and Discussion}

\subsection{Ligand exchange and film morphology}

    \begin{figure}[!htb]
        \centering
            \includegraphics[width=1\textwidth]{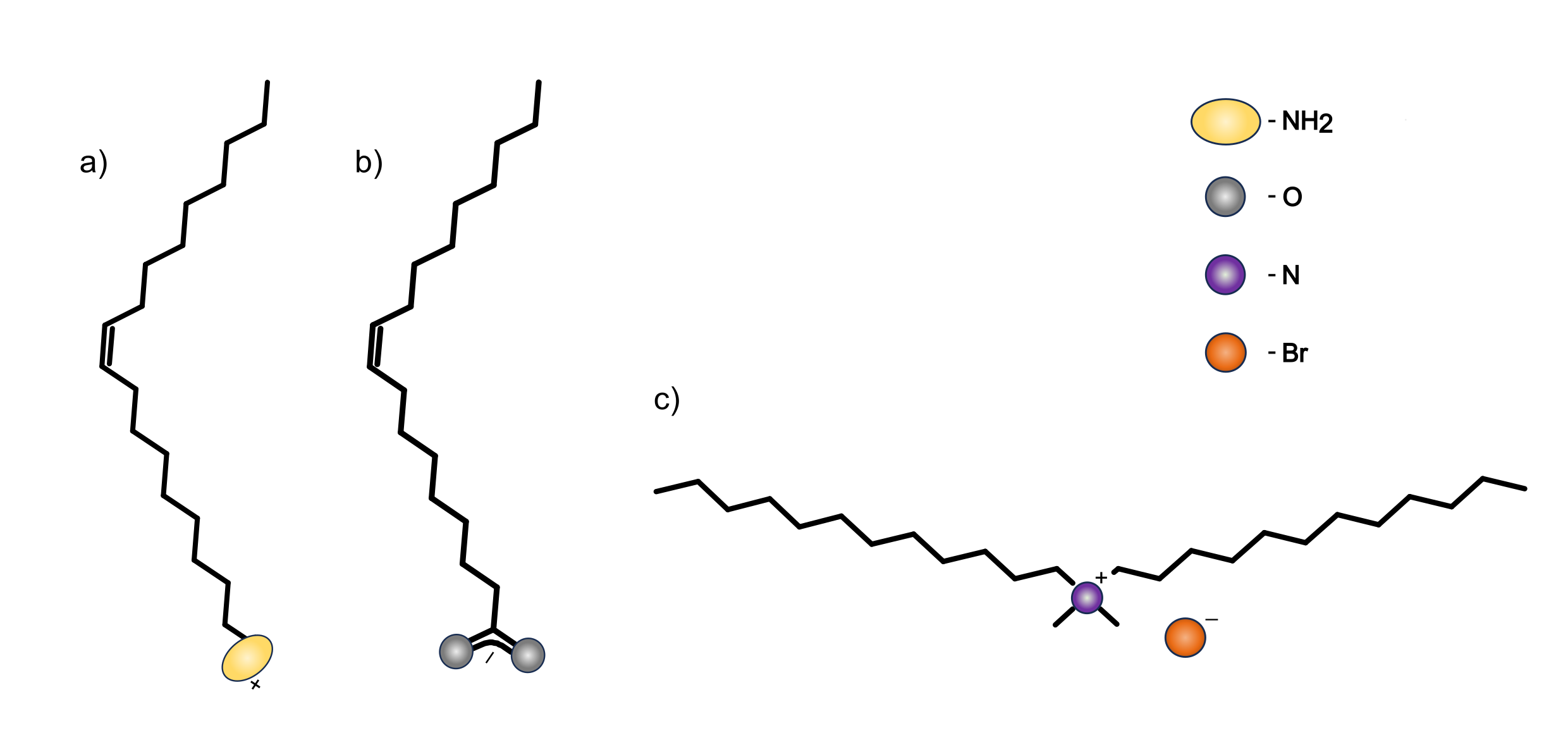}
            \caption{Schematic of a) oleyl amine b) oleic acid and c) didodecylammonium bromide (DDABr)}
        \label{Ligands}
    \end{figure}

    The as-received native colloidal NC suspensions incorporate X-type OA and L-type OAm ligands (Fig.~\ref{Ligands}a \& b) as passivating molecules\cite{Xue2020} . As mentioned earlier, although long alkyl ligands are essential for colloidal stability, they are not favorable for charge injection. Despite various proposed mechanisms on the binding of OA and OAm ligands, it is established that they bind only weakly and are highly dynamic\cite{Fuiza2023}. As a result, these ligands can easily detach from the surface, compromising colloidal stability. In contrast, DDABr  (Fig.~\ref{Ligands}c), with its shorter alkyl chains, significantly improves colloidal stability by simultaneously exchanging anionic and cationic sites on the NC surface\cite{Zaccaria2022,Fuiza2023}. A detailed description on ligand exchange and film fabrication processes is provided in the experimental section.

            \begin{figure}[!htb]
                 \centering
                    \includegraphics[width=1\textwidth]{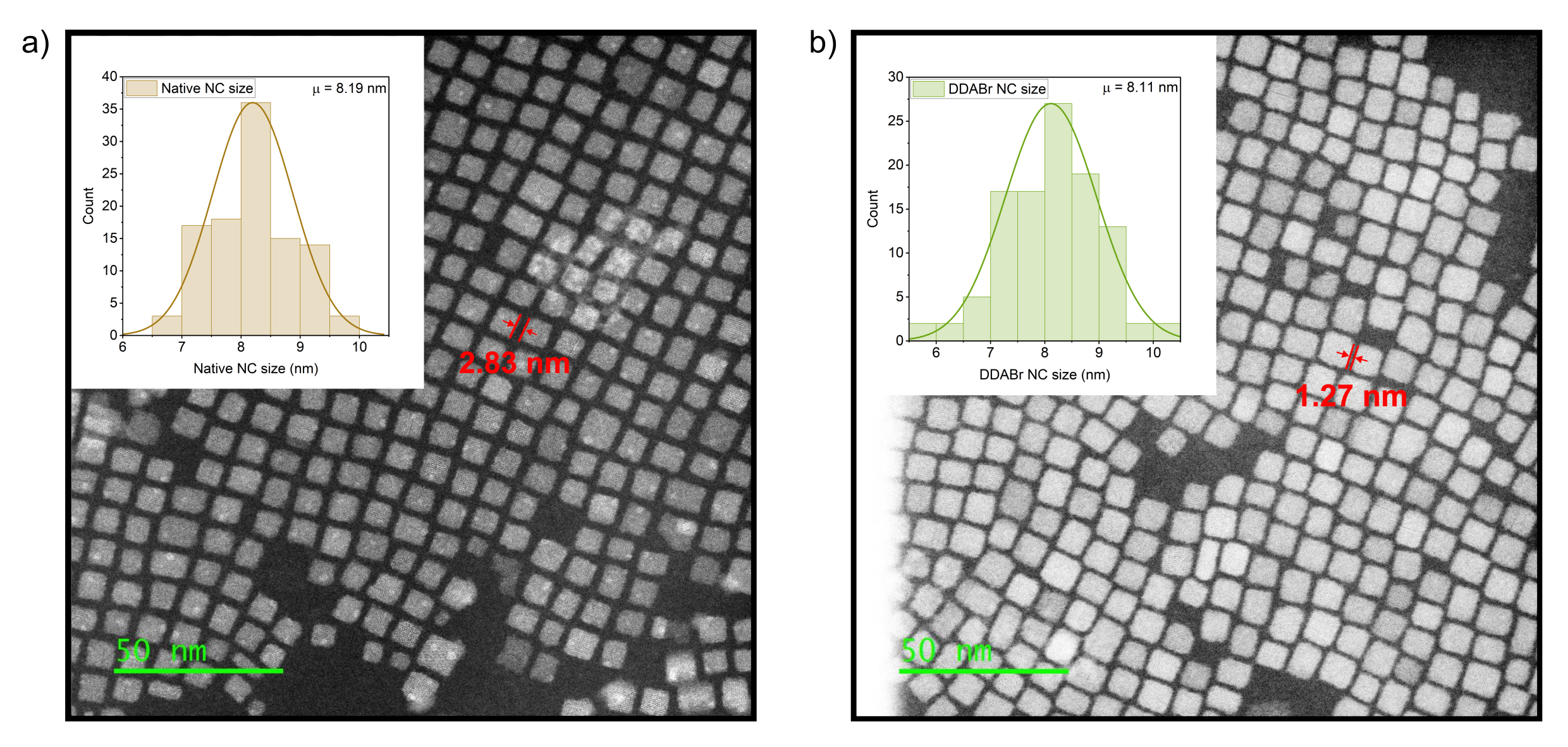}
                     \caption{TEM image of a) OA/OAm covered NCs and b) DDABr covered NCs  with size distribution as inset.}
                 \label{TEM}
             \end{figure}

             Transmission electron microscopy (TEM) images of both the native and ligand-exchanged NCs are presented in Fig.~\ref{TEM}a \& b. Although some etching of the NCs during ligand exchange was expected\cite{Zaccaria2022,Tian2024}, these images show that the size distribution of the ligand-exchanged NCs remains largely unchanged compared to the native system, with an average NC size of approximately 8 nm in both cases. However, the inter-particle distance in the ligand-exchanged system is more than halved relative to the native NCs, from about 2.8 nm to 1.3 nm. This reduction in inter-particle distance could be attributed to two factors. First, the ligand exchange process includes intermittent washing steps which remove a significant portion of the surface ligands, resulting in partial coverage by DDABr only (see Experimental section for further information). Second, the two aliphatic tails of the DDABr ligands may adopt a conformation in which they lie nearly parallel to the NC surface, further reducing inter-particle spacing. This reduced spacing enhances electronic coupling between NCs, which is advantageous for charge transport, however, in the next section, we shall discuss that this improvement mostly affects the positive carriers, i.e., holes only.

            \begin{figure}[!htb]
                 \centering
                    \includegraphics[width=1\textwidth]{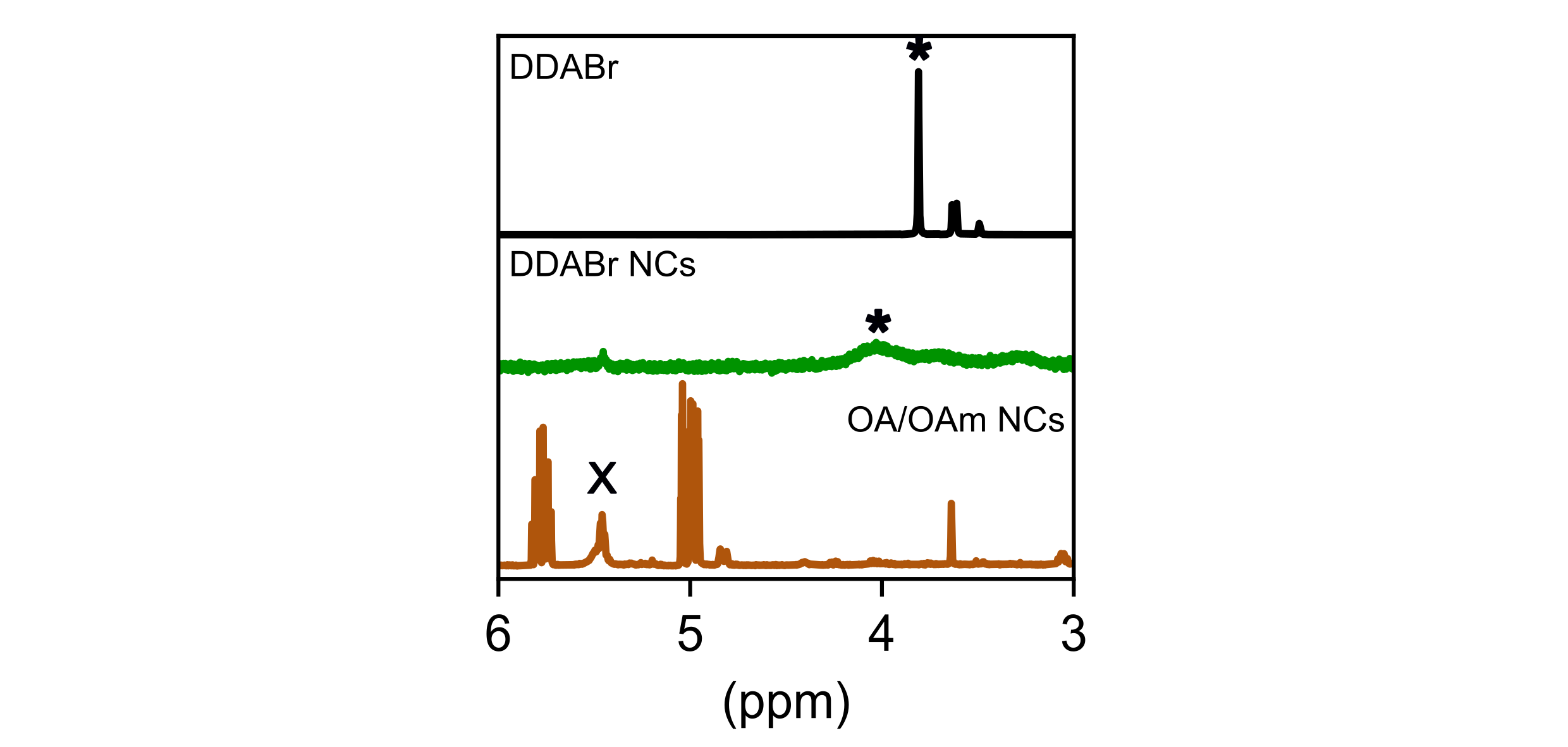}
                    \caption{1H NMR spectra (in toluene-d8) of DDABr as reference (black), ligand exchanged DDABr-covered NCs (green) and native OA/OAm NCs (brown).}
                \label{NMR}
            \end{figure}

            To confirm the exchange of the organic ligand shell, \textsuperscript{1}H NMR spectroscopy was performed. Fig.~\ref{NMR} shows a comparison of the spectra of native NCs covered with OA/OAm, DDABr-covered NCs and DDABr alone in toluene-d\textsubscript{8}. The spectra display the disappearance of the alkene protons of OA/OAm at approximately 5.5 ppm (marked by $\times$) for DDABr NCs. In addition, a signal at 4 ppm arises in the exchanged sample, which can be assigned to the protons at the methyl groups of DDABr (marked by *). This signal shows a peak broadening and a downfield shift which is consistent with a DDABr species bound to NCs\cite{Zaccaria2022,De-Roo2016}. Both of these determined signals indicate a successful ligand exchange from OA/OAm to DDABr. Further investigations by quantitative-NMR (qNMR) spectroscopy (see Fig. \ref{qNMR}, Supplementary Information) on the ligand coverage indicate about 226 ligands/NC for DDABr NCs after two washing steps.
            

\subsection{Electronic properties}

    \subsubsection{Spectro-electrochemistry}

        \begin{figure}[!htb]
            \centering
                \includegraphics[width=1\textwidth]{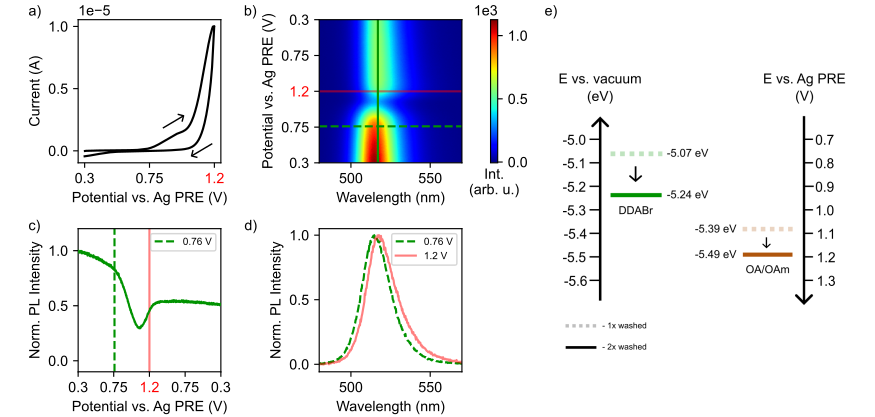}
                \caption{Exemplary spectro-electrochemistry measurement on DDABr-capped CsPbBr\textsubscript{3} NCs (after one washing step). a) Cyclic voltammetry scan measured at a scan speed of 2mV/s in PC/TBAHFP (0.1 M) with arrows indicating the scan direction. b) Potential-dependent PL spectra with marked onset potentials of the oxidation (green dashed line) and the switching potential (red solid line). c) Normalized potential-dependent PL intensity at the emission peak extracted from b). d) Spectra slices at the oxidation potential and switching potential extracted from b). e) Graphical summary of the results for native and DDABr capped NCs after one and two washing steps.}
            \label{SEC}
        \end{figure}

            One way to investigate the absolute band edge positions of NCs is by spectro-electrochemistry (SEC)\cite{Samu2018,Mulder2021,Ashokan2021}. The basic idea is that by electrochemical charge injection into the NCs, their photoluminescence (PL) intensity decreases, which can be monitored simultaneously by optical spectroscopy. In this work, the valence band (VB) position of OA/OAm and DDABr capped NCs has been studied by SEC in the oxidative regime. Additionally, the effect of varying ligand coverage on the NC surface has been investigated, since ligand stripping was reported to improve the performance of NC LEDs\cite{Dai2024}. 
            
            An exemplary measurement of DDABr capped NCs after one washing step is shown in Fig. \ref{SEC}a-d. SEC-PL measurements are conducted with a potential scan of 2 mV/s in a propylene carbonate/tetrabutylammonium hexafluorophosphate solution (PC/TBAHFP) against a Ag pseudo-reference electrode (Ag PRE). Before applying a potential to the samples, photobleaching of the sample due to laser power is determined, with detailed information in the SI (Fig. \ref{PL no bia}). Cyclic voltammetry in the oxidative direction (Fig. \ref{SEC}a) displays a current increase starting at 0.75 V, indicating the oxidation of NCs, ie., electron  transfer from the NC to the working electrode, followed by a potential sweep reversing at 1.2 V. The PL spectra is recorded under applied bias in steps of 2 mV, and is depicted as a heat map in Fig. \ref{SEC}b. From the PL intensity over the applied bias (Fig. \ref{SEC}c), a PL decay at the oxidation potential is observed. Thereafter, a partial recovery of PL is observed as the potential returns to the intial value. Spectral slices at the oxidation onset and the switching potential reveal a slight red shift of the oxidised NCs (Fig. \ref{SEC}d). These measurements were also performed on twice washed DDABr capped NCs and the native, i.e., OA/OAm capped ligands, also for single and doubly washed samples. (Fig. \ref{OA-once-precipitated}-\ref{DDABr-twice-precipitated}). 
            
            Finally, Fig. \ref{SEC}e summarizes the oxidation potentials determined for the four samples referenced to a ferrocene/ferrocenium redox couple (Fig. \ref{Reference measurement}) and to the vacuum scale. The obtained values are in good agreement for the VB edge position in the range of -5.07 eV to -5.5 eV vs. vacuum, in comparison with previous work by Mulder et al\cite{Mulder2021}. Overall, two trends can be deduced from this data. Firstly, washing leads to a decrease in the VB onset potential by approximately 0.17 eV for DDABr-exchanged NCs and 0.1 eV for native NCs. Secondly, DDABr shifts the VB edge by approximately 0.3 eV towards the vacuum level compared to OA/OAm at similar ligand densities. It is to be noted that the DDABr twice-washed samples (with their VB energy at -5.24\,eV relative to the vacuum level) have been used further on throughout the remainder of this work.

    \subsubsection{Photoelectron spectroscopy}
            
        \begin{figure}[!htb]
                \centering
                 \includegraphics[width=1\textwidth]{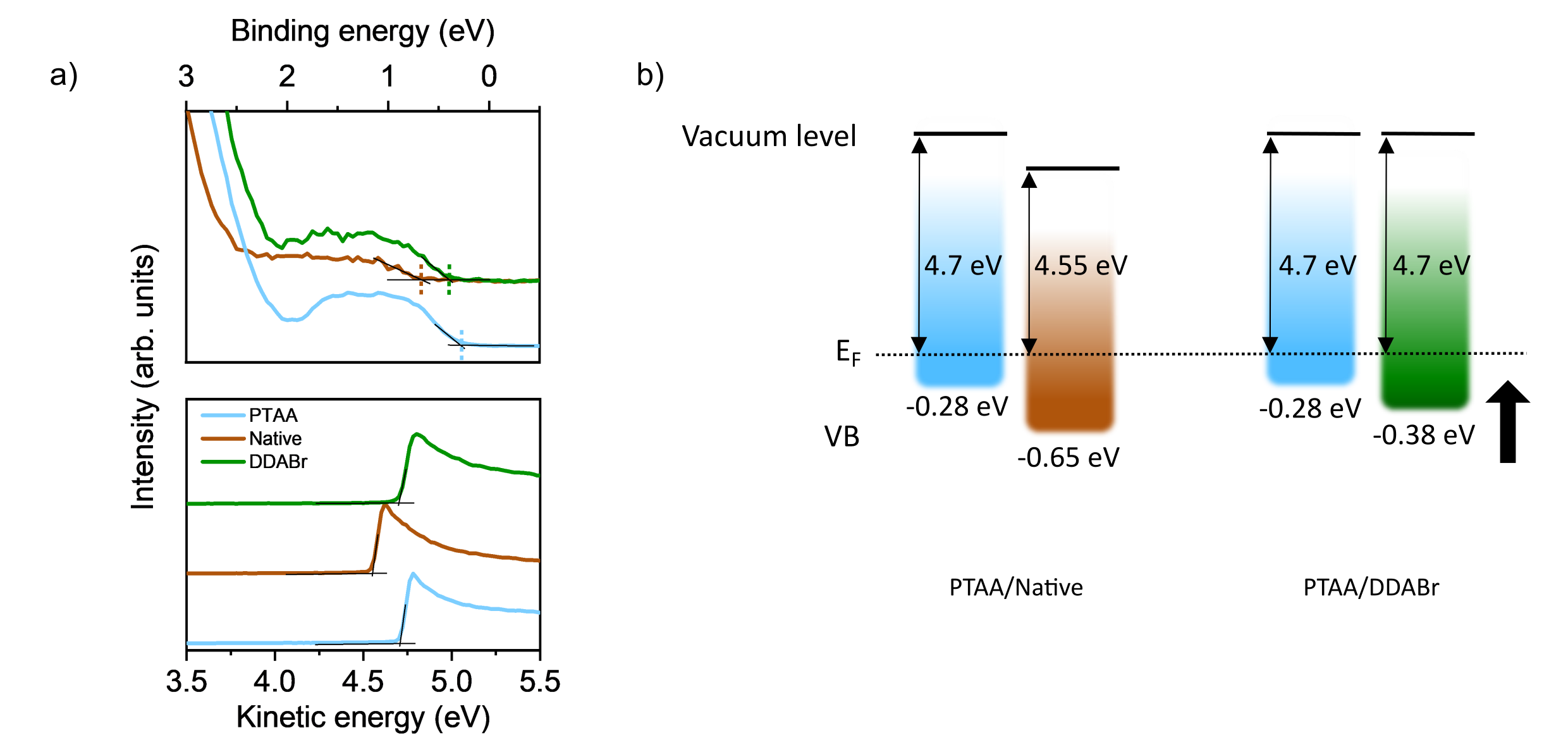}
                 \caption{UPS data on native (brown) and DDABr capped NCs (green) spincoated on PTAA substrates (blue). a) Valence band (top) and secondary electron cut-off (bottom) spectra. b) Schematic representation of energy alignment obtained from UPS data.}
                 \label{PES data}
        \end{figure}

        To gain deeper insight into the energy levels of CsPbBr\textsubscript{3} NCs before and after ligand exchange, photoelectron spectroscopy (PES) was conducted on thin films used in PeLEDs. NC films were deposited on ITO/PEDOT:PSS substrates coated with PTAA \sloppy(Poly[bis(4-phenyl)(2,4,6-trimethylphenyl)amine]), which serves as the hole transport layer (HTL) in PeLEDs. Ultraviolet photoelectron spectroscopy (UPS) was used to determine the energy level alignment at the PTAA/NC interface. As shown in Fig. \ref{PES data}a (upper panel), the VB onset appears at 0.28 eV for PTAA, 0.65 eV for native NCs and 0.38 eV for DDABr-capped NCs. The secondary electron cutoff, i.e., the work function (Fig. \ref{PES data}a, lower panel) indicates a slight vacuum level shift for native NCs. The corresponding energy level alignment depicted in Fig. \ref{PES data}b and reveals a significantly reduced hole injection barrier of only about 0.1\,eV from PTAA into the NCs for DDABr ligands, while the barrier is almost 0.4 eV for native NCs.

        X-ray photoelectron spectroscopy (XPS) further confirmed the chemical composition of both NC systems. Survey spectra (Fig. \ref{XPS survey}) show an increased photoelectron signal from perovskite core levels (Cs 3d, Pb 4f, Br 3d) relative to C 1s in DDABr NCs, indicating successful ligand exchange. Quantitative XPS analysis (Fig. \ref{XPS elements}, Tab. \ref{RAR}, Supplementary Information) reveals a Br/Pb ratio increase from 1.85 (native NCs) to 2.57 (DDABr NCs), suggesting halide vacancies in native NCs, which are subsequently filled by Br\textsuperscript{-} ions during ligand exchange. These vacancies introduce n-type defect states\cite{Cohen2019}, which likely contribute to the observed energy-level shifts.

    \subsubsection{Single-carrier devices}

        \begin{figure}[!htb]
            \centering
                \includegraphics[width=1\textwidth]{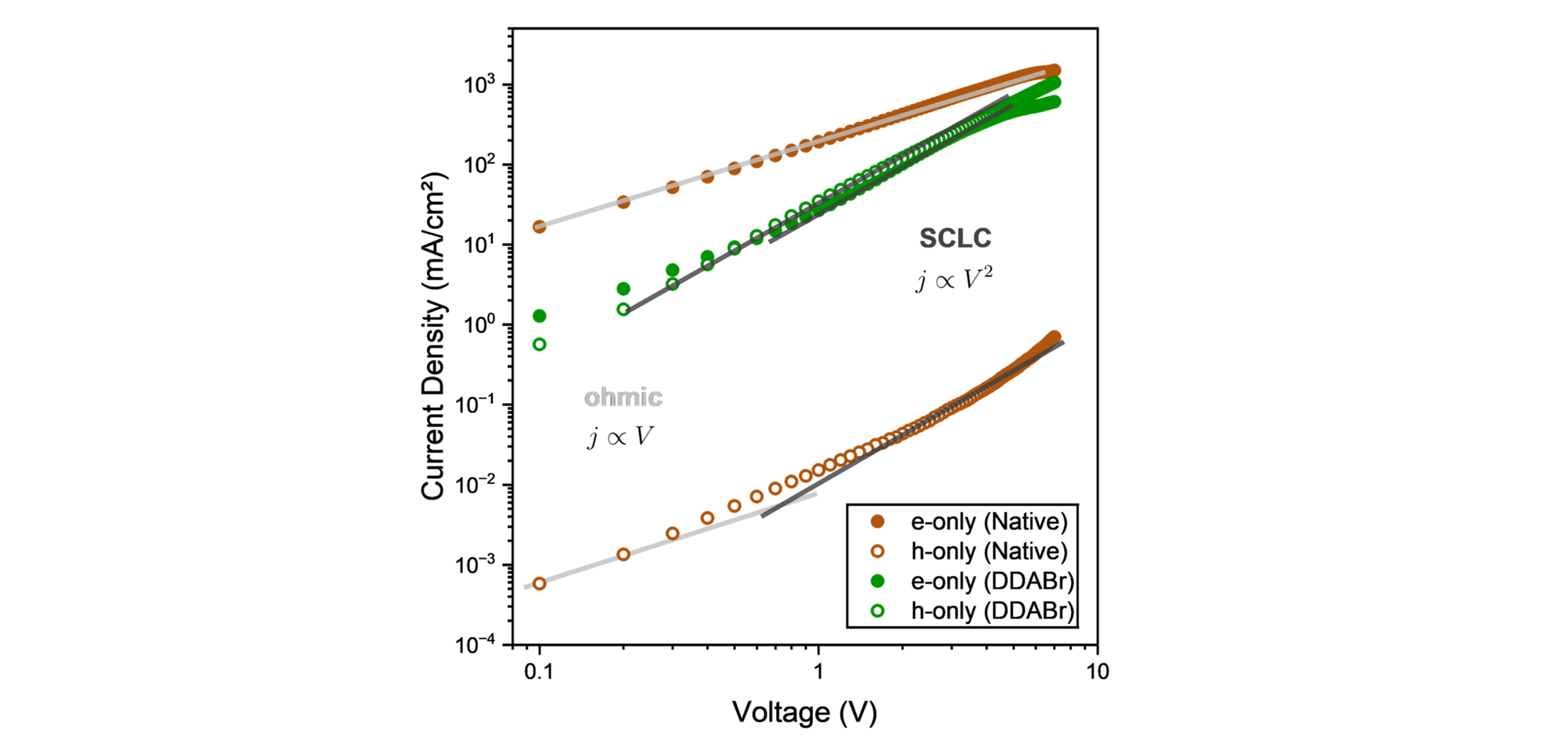}
                \caption{Current density vs voltage characteristics of single-carrier hole-only (hollow circles) and electron-only (filled circles) devices with native and DDABr-capped NCs.}
            \label{Single carrier}
        \end{figure}

            A persistent challenge in PeLEDs that have native NCs as emitter system is the imbalance in carrier transport, particularly, poor hole transport\cite{Tassilo2022}. In this regard, single-carrier devices were fabricated to assess the hole and electron transport in both the native and ligand-exchanged NC systems. Figure \ref{Single carrier} illustrates the hole-only (hollow circles) and electron-only (filled circles) devices for the native and ligand exchanged systems. The analysis uses simple power-law dependencies of current density ($j$) on the applied voltage ($V$), which are characteristic for different transport regimes \cite{A-k2015}.
            
            A first examination of the native system (brown color) reveals a high electron current with an ohmic dependence on voltage, indicating low resistance in electron conduction through the native system. In contrast, the hole current is extremely low, about four orders of magnitude lower than the electron current, and follows an almost quadratic dependence on voltage, which is typical for space charge-limited conduction (SCLC) according to the relation, $j_\mathrm{SCLC} = (9/8)\,\varepsilon \varepsilon_0\, \mu \times (V^2/d^3)$. With a NC layer thickness of $d=$\,30 nm and a dielectric constant of $\varepsilon=7$\cite{Futscher2020}, the hole mobility $\mu$ was determined to be approximately 6.5 $ \times$ 10\textsuperscript{-10} cm\textsuperscript{2}/Vs. Note that due to the absence of SCLC, electron mobility could not be reliably estimated for the native NC system. However, DDABr-capped NCs exhibit a more balanced carrier transport (green color), with a significantly improved hole current compared to the native system. Based on the observed SCLC behavior, the hole and electron mobilities are estimated at approximately 2.1 × 10\textsuperscript{-6} cm\textsuperscript{2}/Vs and 1.5 × 10\textsuperscript{-6} cm\textsuperscript{2}/Vs, respectively. Notably, the hole mobility increases by almost four orders of magnitude after ligand exchange, providing strong evidence of enhanced hole injection and transport in DDABr-exchanged NCs.

 \subsection{DFT calculations}
         
\begin{figure}[!htb]
        \centering
            \includegraphics[width=1\textwidth]{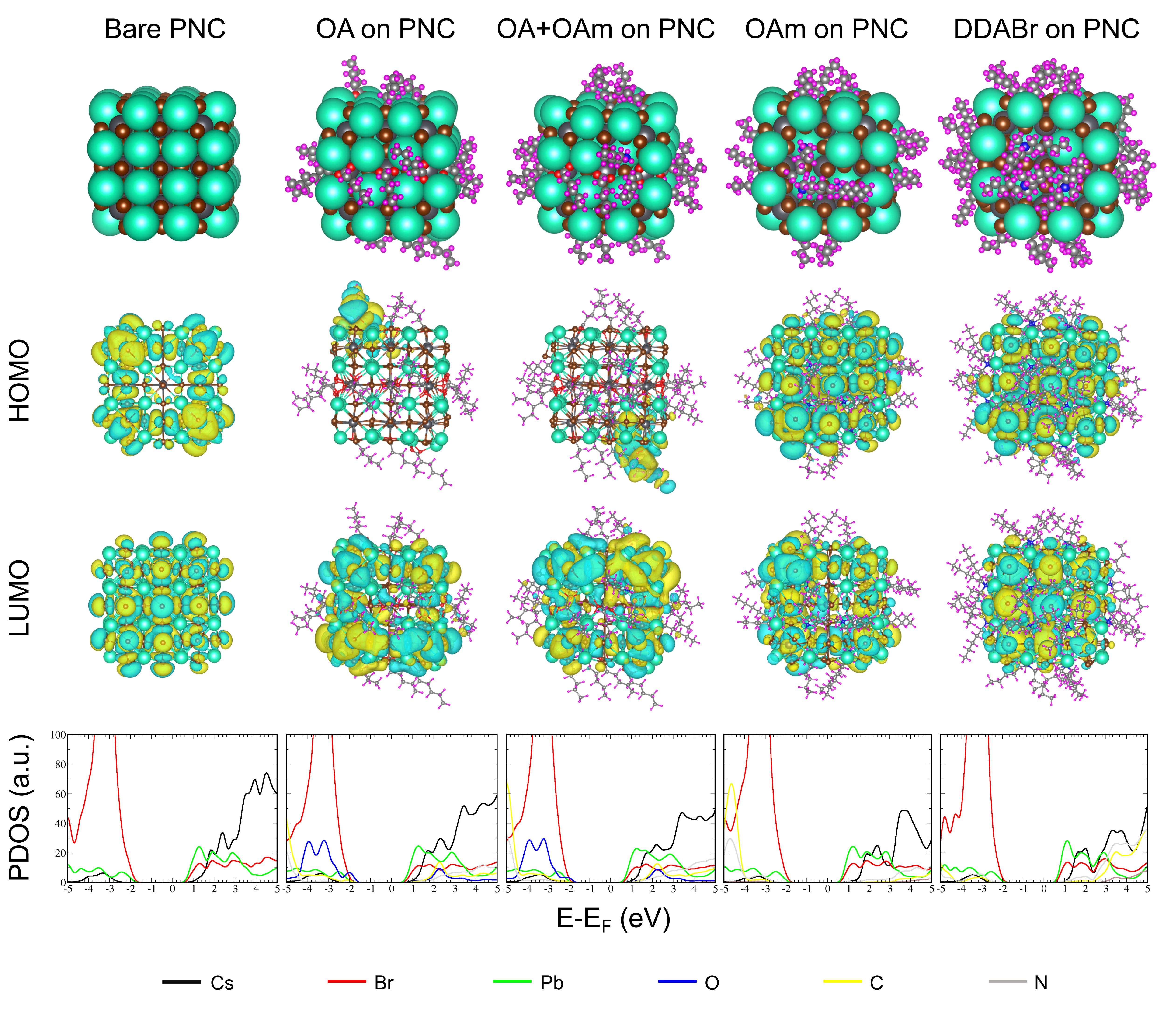}
            \caption{Schematic representation of the bare NC along with the various ligands attached to the surface. Top row from left to right represent the space-filling model of the bare NC, a NC with 18 hexanoates (mimicking OA), a NC with 18 OA and one hexylammonium (mimicking OAm), a NC with 18 OAm and, finally, a NC with 18 dihexylammonium (mimicking DDABr). Second and third rows from left to right represents the PBE-D3 computed HOMO and LUMO, respectively, of the corresponding species shown at the top. Color code: Cs- light green; Pb- dark grey; Br- brown; O- red, N- blue; H-light pink and C- light grey. Bottom row from left to right represents the PDOS calculated from the frontier molecular orbitals.}
        \label{DFT}
    \end{figure}

            DFT calculations were carried out using the CP2K software package\cite{Kuehne2020} (version 9.1) to understand the electronic structure of native NCs and the effect of the various ligands on the frontier molecular orbitals, i.e. the highest occupied molecular orbital (HOMO) and the lowest unoccupied molecular orbital (LUMO), VB and conduction band (CB), respectively. First, a neutral cubic model of CsPbBr\textsubscript{3} containing 189 atoms (Cs\textsubscript{54}Pb\textsubscript{27}Br\textsubscript{108}) possessing a dimension of 1.7 nm was chosen with the Cs-Br terminated surface or facet to begin the calculation on a bare NC\cite{Bodnarchuk2019} -- see Fig.\ref{DFT}. Numerous experimental and theoretical reports support the fact that the Cs-Br terminated surface is the commonly exposed surface rather than the Pb-Br facet\cite{Lapkin2022,Wahl2022,Bodnarchuk2019,Wei2016,Ten2016} . Furthermore, the anion/lead ratio (Br/Pb) was fixed at 4.0 for this composition of the NC, which is the upper bound and somewhat overestimated compared to the commonly encountered ratio of 2.7-3.3\cite{Bodnarchuk2019}. At this geometry and composition, the bare NC exhibited a band gap of 2.69 eV (also denoted as HOMO-LUMO gap in the following), which is a little larger than the experimental value of 2.3 eV. This discrepancy can be attributed to the size chosen for the bare PNC, (which was minimized to reduce computational cost), as well as the use of the PBE functional in DFT calculations, which inherently involves a trade-off, balancing the accuracy of the HOMO-LUMO gap estimation. This has also been encountered in previous reports on electronic structure determination of cubic NCs\cite{Lapkin2022,Wahl2022,Bodnarchuk2019,Ten2016}. As expected, the VB (or HOMO) states are dominated by the halide (Br) 4p-orbitals and the CB (or LUMO) states consist of virtual or anti-bonding p-orbitals from both Pb and Br atoms (see Fig.~\ref{DFT}, bottom row, PDOS). 

            In the next step, the ligands under study, namely, OA, OAm and DDABr were successively attached at the PNC surface in all six directions ($\pm x$, $\pm y$ and $\pm z$) to understand their effect on the frontier molecular orbitals and the density of states. To reduce the computational cost and allow for efficient attachment of the ligands onto the PNC surface, the large chain lengths were reduced to six-numbered carbon-containing ligands for all the cases\cite{Wahl2022}. hexanoate ligands were added to the Cs-Br facet of the PNC successively by removing bromide anions one by one from each side of the PNC. In every case, the charge neutrality of the entire system was maintained. In the case of oleate ligands, irrespective of their density around the PNC, ligand-centered HOMO or localized states were observed, which are also seen as trap states in the valence region of the partial density of states (PDOS)  (Fig. \ref{DFT}, OA on PNC). Depicted as blue curves on the PDOS plot, these trap states are dominated by the carboxylic oxygen-centered p- and/or $\pi$-orbitals. These trap states are present even when a combination of OA and OAm ligands are attached to the Cs-Br surfaces of the PNC (Fig. \ref{DFT}, OA + OAm on PNC). However, when the ligands that are attached to the surface are either OAm only or DDABr ligands, no ligand-centered trap states were observed in the PDOS of the valence region and the HOMO is delocalized over the entire NC (Fig. \ref{DFT}, OAm on PNC / DDABr on NC).  

    \begin{figure}[!htb]
         \centering
            \includegraphics[width=1\textwidth]{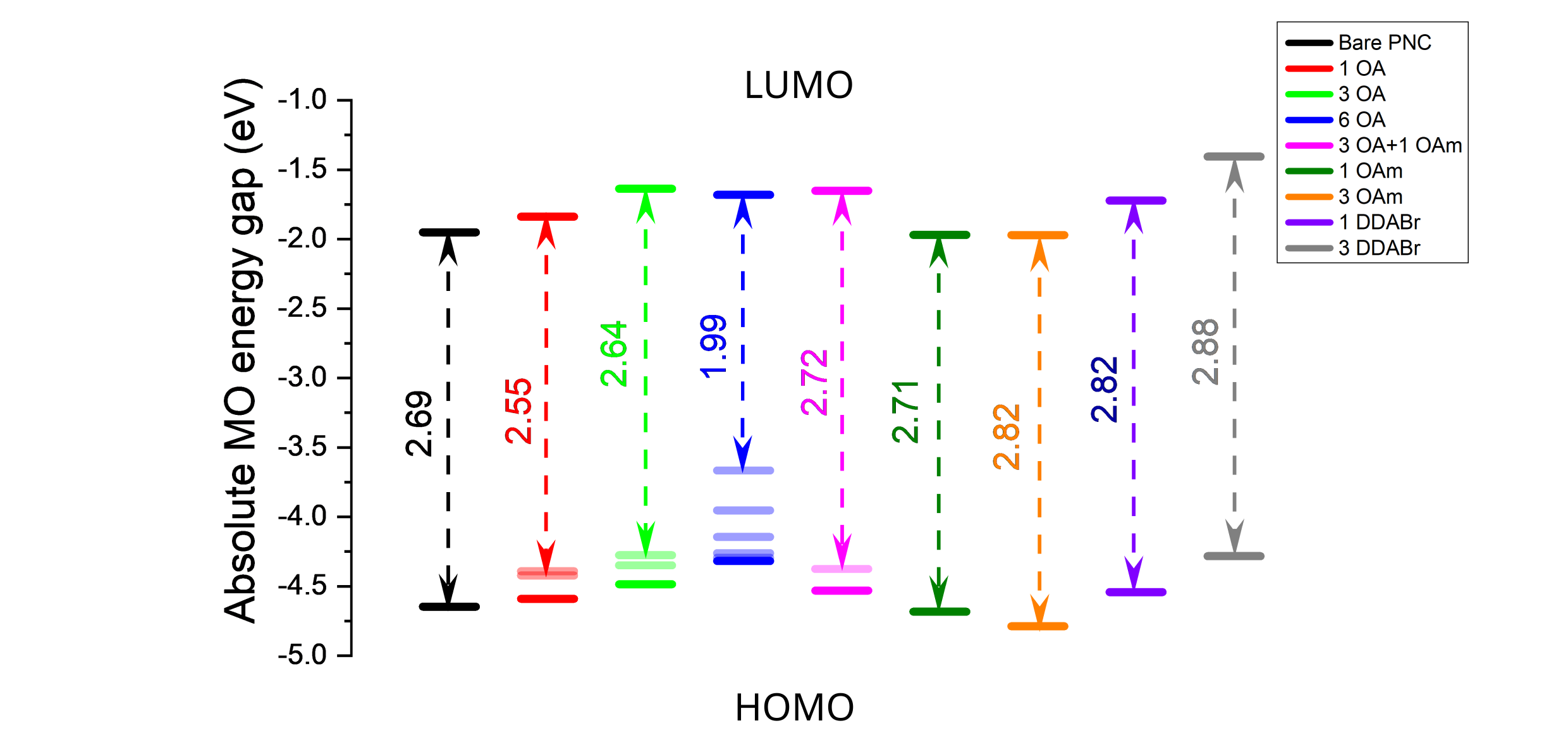}
            \caption{PBE-D3 computed frontier molecular orbital energies of PNCs bonded with the various quantities of ligands per facet. The transparent levels represent the ligand-centric hole trap states. OA - Oleate (hexanoate); OAm - oleylammonium (hexylammonium); DDABr - di-hexylammonium ligands. }
        \label{Energy levels}
    \end{figure}

            Furthermore, DFT calculations have also been performed at moderate and excess concentrations of ligands. The obtained absolute energies of the frontier molecular orbitals are plotted in Fig. \ref{Energy levels}. It can be seen that oleates or, particularly, the carboxylate oxygen create hole trap states in between the usual Br and Pb-centred HOMOs and LUMOs. On the contrary, this trap (or ligand-centric HOMO) is absent in the case of OAm or DDABr ligands. Notably, the energy gap increases when these ligands are attached to the NC. Very importantly, in the case of the DDABr, the HOMO is lifted in energy compared to the OA/OAm bound PNCs, in agreement with the SEC and UPS measurements. In order to investigate the origin of trap states, the nature of the binding mode of oleate ligands to the NC surface was examined. A similar kind of study\cite{Voznyy2011} has already been performed on CdSe QDs, where oleate  binding on different facets of CdSe QDs could result in the occurrence of trap states\cite{Voznyy2011,Cossedu2023}. By contrast, NCs are cubic and therefore have equivalent facets on all sides. The only difference that could arise is to which atom the oleate binds, i.e., Pb\textsuperscript{2+} or Cs\textsuperscript{+}, which depends on the termination of the surface and defects. To generate models for binding the anionic ligands such as oleate to the NC surface, anion vacancies were created by removing bromide anions from the surface. Here, two different scenarios were considered that can both represent the oleate passivated surface but, concurrently, show two completely different binding modes. In Fig. \ref{One oleate HOMO}, the two different modes of binding of oleate ligands are shown. In one case, the two oxygen atoms of a carboxylate group are bonded to Cs\textsuperscript{+} ions (Fig. \ref{One oleate HOMO}a \& b, whereas in the other case, all the carboxylate oxygen(s) are directly bonded to a Pb\textsuperscript{2+} ion (Fig. \ref{One oleate HOMO}c \& d. It is to be noted here that both possibilities generate a charge-neutral cluster model in order to accommodate the oleate group. Interestingly, these two scenarios produce quite different natures of HOMOs. In one case, the HOMO is ligand or oleate centric, whereas, in the other case, the HOMO consists of the usual bromide-based molecular orbitals. The former possibility is more likely to occur experimentally as it is already known that the Cs-Br surface termination is more probable as the NC size increases\cite{Bodnarchuk2019}. Since commercial NCs have been used in this work, it is difficult to comment on the amount of ligands on the surface and the exact composition of the NCs. However, it is clear that the presence of OA impedes carrier injection into the NC as well as inter-particle transport, in particular, for holes.

\subsection{PeLEDs}

    \begin{figure}[!htb]
        \centering
            \includegraphics[width=1\textwidth]{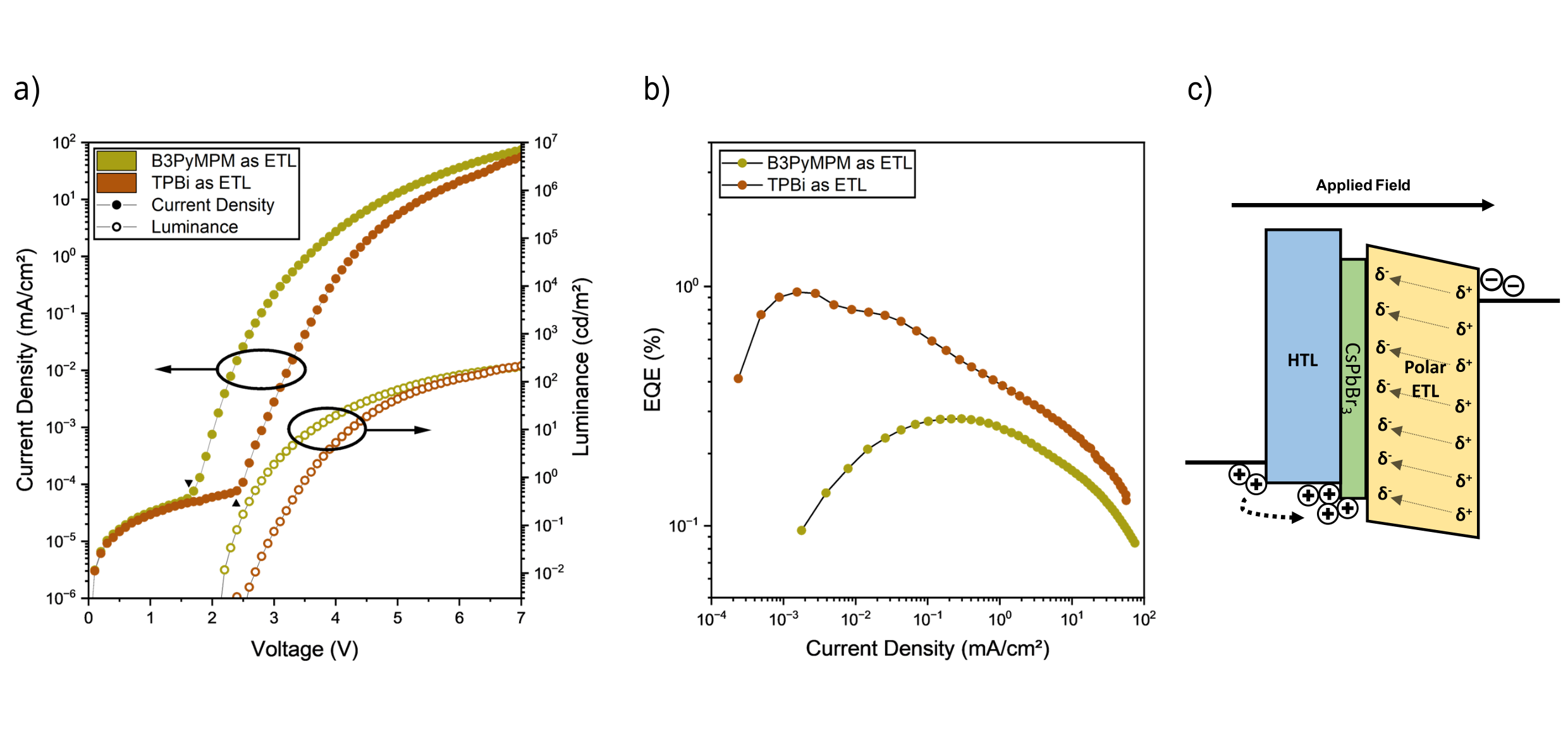}
            \caption{a) JVL and b) EQE vs current density plots of native PeLEDs comparing the performance of TPBi (polar) and B3PyMPM (non-polar) as ETLs. c) Schematic representation of hole accumulation before V\textsubscript{on} due to GSP by the polar ETL. }
        \label{native LED}
    \end{figure}

        A major challenge in achieving efficient LEDs based on native CsPbBr\textsubscript{3} NCs is their inefficient hole injection, resulting in imbalanced carrier transport and recombination, and subsequently excited-state quenching by Auger-like recombination. This leads to low efficiency and pronounced efficiency roll-off at higher currents, as demonstrated in our previous study on doped NCs\cite{Tassilo2022}. Based on our findings so far, ligand exchange with DDABr is expected to mitigate this issue by reducing the hole injection barrier and achieving balanced carrier mobility. However, successful LED fabrication also requires appropriate (organic) transport layers. As a first step, PTAA was chosen as the HTL due to its superior hole mobility compared to PVK [Poly(9-vinylcarbazole)]. Additionally, the role of the ETL must be considered, as it not only facilitates electron transport and injection into the NCs but also functions as an optical spacer, enhancing light outcoupling, similar to its role in organic LEDs\cite{Bruetting2013}.
     
        Commonly used materials in PeLEDs include B3PyMPM (4,6-Bis(3,5-di(pyridin-3-yl)phenyl)-2-methylpyrimidine) and TPBi \sloppy[2,2',2''-(1,3,5-benzinetriyl)-tris(1-phenyl-1-H-benzimidazole)], to name two  prototypical materials. While these materials exhibit different energy levels and electron mobilities, their most significant difference is often overlooked in the context of PeLEDs. TPBi, owing to its notable permanent dipole moment, tends to exhibit SOP upon evaporation, which in turn induces a non-vanishing electric field across the organic film, also known as giant surface potential (GSP)\cite{Noguchi2022,Hofmann2025}. The direction of this field hampers electron flow through the ETL at low applied voltages, while simultaneously promoting hole injection and accumulation at the HTL/emitter interface even before the device reaches its turn-on voltage.

        Fig. \ref{native LED}a illustrates the current (density)-voltage-luminance (jVL) characteristics of native NC PeLEDs employing B3PyMPM, a non-polar ETL exhibiting almost no GSP and TPBi as polar ETL. Details on the LED fabrication process can be found in the experimental section. The native PeLED with the non-polar ETL (yellow) exhibits a lower turn-on voltage (V\textsubscript{on}) at about 1.7 V, however, the luminance onset is at a higher voltage of about 2.2 V. This is an indication of imbalanced carrier injection, i.e. the current is dominated by one carrier type, while luminance sets in later when the second carrier type is injected. Correspondingly, the external quantum efficiency (EQE) shown in Fig. \ref{native LED}b is rather low (in the 10\textsuperscript{-1} range) and exhibits a strong dependence on the current density itself with a maximum in the range of $10^{-1}-10^{-2}$\,mA/cm$^2$. Such carrier density dependent behavior is typical for a competition of different processes, like trap-assisted recombination, bimolecular recombination and Auger quenching\cite{Deschler2016}.
     
        In the case of native PeLED employing the polar ETL (brown), an obvious shift in the current onset (V\textsubscript{on}) to about 2.4 V; however, the luminance onset is only slightly higher at 2.7 V. The higher V\textsubscript{on} arises from the GSP of the polar ETL, which has to be overcome before electrons can flow through the ETL. The GSP value of TPBi on native NCs is about 112 mV/nm (see Fig. \ref{KP}), which is two times higher than the reported values in organic devices\cite{Hofmann2025}. Importantly, the native TPBi PeLED reaches a significantly higher EQE of almost 1\%, specifically at very low current density of $10^{-3}$\,mA/cm$^2$. This is again a direct consequence of the GSP in the TPBi ETL, which leads to hole accumulation in the HTL prior to reaching the turn-on voltage (Fig. \ref{native LED}c, similar to OLEDs, where this process is well established\cite{Noguchi2022}. This means that carrier balance is improved by this sub-turn-on hole accumulation, but with increasing current this advantage disappears and the EQE of the native TPBi device also drops significantly. Finally, both devices (native B3PyMPM and TPBi) reach the same maximum luminance of about 200 cd/m² at 7 V. 

        In order to improve charge injection, the native NCs were washed with ethyl acetate once and LEDs utilizing TPBi have been fabricated (Fig. \ref{Washed vs unwashed}). These LEDs exhibited a higher current density in comparison to unwashed NCs with slightly improved luminance of about 400 cd/m² at 7 V. However, they demonstrated a similar trend with EQE against current density with a peak EQE at the beginning of about 2.3\%, followed by a roll-off at higher current densities. Despite the improved EQE achieved with TPBi as the ETL, the native NC PeLEDs continue to face challenges, including a relatively large hole injection barrier between the PTAA HTL and the native NCs (as evidenced by the SEC and UPS data presented above). Additionally, hole trap formation as a result of the binding of OA ligands to the NC surface remains an issue (as indicated by the DFT calculations).

    \begin{figure}[!htb]
        \centering
            \includegraphics[width=0.9\textwidth]{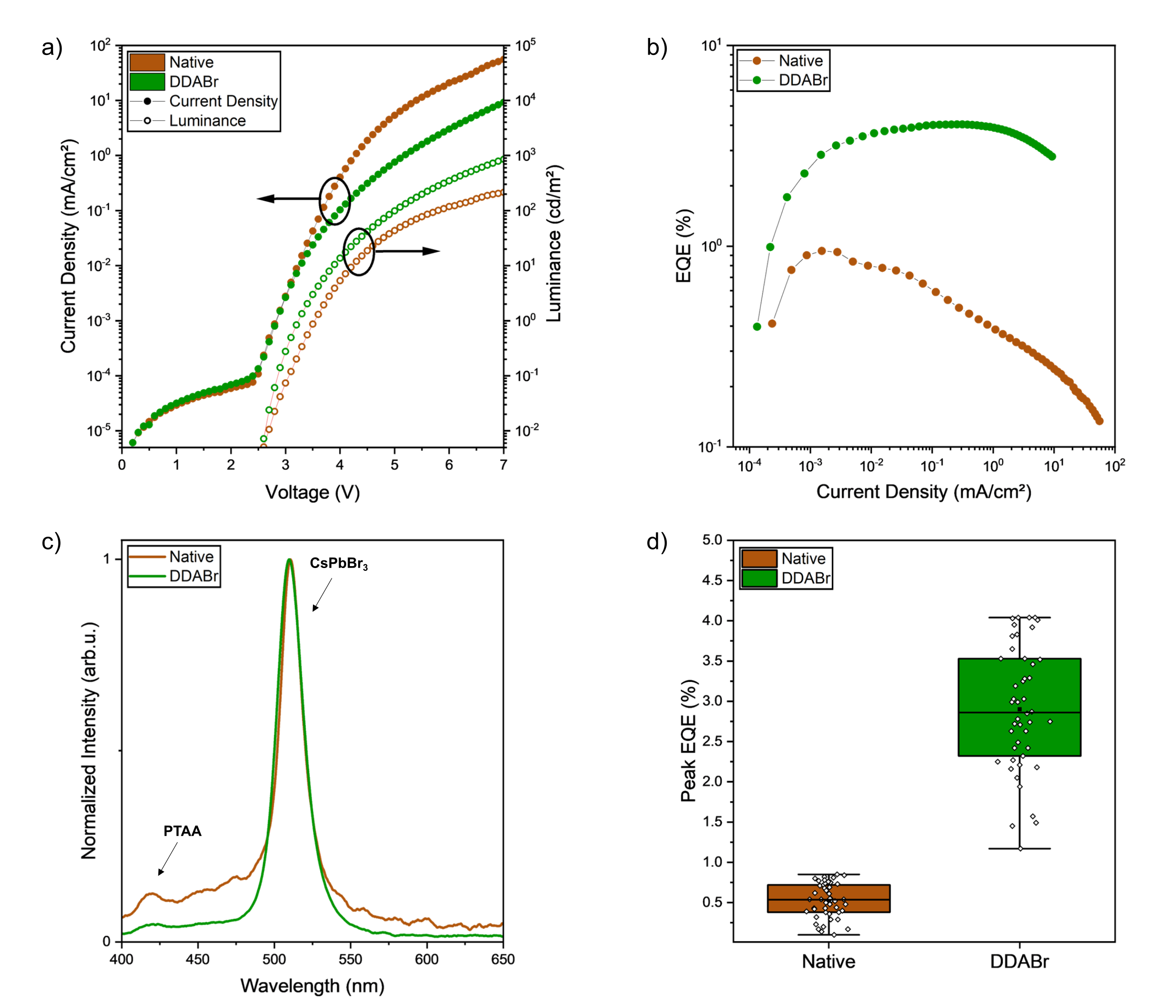}
            \caption{a) JVL curves b) EQE vs current density, c) EL spectra and d) reproducibility statistics of native and DDABr PeLEDs.}
        \label{LED}
     \end{figure}

        To address these concerns, we proceeded to fabricate PeLEDs using NCs with DDABr-ligand exchange and the polar TPBi ETL. Figure \ref{LED}a presents the jVL characteristics of native and DDABr-ligand exchanged NC PeLEDs in comparison. DDABr NC LEDs demonstrate similar V\textsubscript{on} as the native NC LEDs at around 2.3 V (with a GSP value of TPBi at 126.8 mV/nm, see Fig. \ref{KP}) and luminance onset at about 2.6 V. However, they reach higher luminance, with peak brightness close to 1000 cd/m² at 7 V, at a reduced current density. This directly translates to an improved EQE (Fig. \ref{LED}b) of up to 4\%. Most importantly, this high EQE is preserved over a broad range of current densities, which is a clear indication to a balanced carrier injection and recombination in the case of DDABr-capped NC PeLEDs. Note that the achieved EQEs are in good agreement with the results published by Dai et al.\cite{Dai2024} on a similar PeLED stack.
     
        Both native and DDABr ligand-exchanged devices exhibit electroluminescence (EL) with a peak emission at 513 nm and a full-width at half-maximum (FWHM) of approximately 17 nm (Fig. \ref{LED}c). However, the native NCs show an additional weak emission centered around 425 nm, which is attributed to some parasitic emission from the PTAA layer. This suggests that in the case of native NCs, the large excess electron current is partially injected into the HTL and recombines with holes in that region, a phenomenon that is significantly reduced in the DDABr-capped NC devices, where the recombination is more localized within the NC layer, as desired. 
     
        Figure \ref{LED}d presents the reproducibility statistics of the ligand exchange process, where the peak EQE of ten samples, with each sample having four pixels, yielded an average (peak) EQE of 3\% for DDABr and about 0.49\% for native NC PeLEDs. Importantly, the devices with DDABr ligands reach their peak EQE at application-relevant current densities of 1\,mA/cm$^2$, while the LEDs with native NCs have their peak at very low currents ( $10^{-3}$\,mA/cm$^2$). These findings highlight the critical role in choosing the right ligand to improve carrier balance in PeLEDs, through the combined effect of better energy level alignment for hole injection and a reduction of hole traps on the NC surface along with the incorporation of a polar ETL, which collectively enhances the performance of PeLEDs.
\section{Conclusion}

The present work highlights the impact of alkyl ligands, specifically OA, OAm and DDABr, on carrier transport and the optoelectronic performance of CsPbBr\textsubscript{3} nanocrystal based PeLEDs. Though obtaining high PLQY in native NCs can be realized through complete passivation of OA/OAm ligands of the surface, this approach is counter-productive with respect to electrical accessibility. Apart from the insulating nature of these alkyl ligands, it is evident from SEC and UPS that the functional groups attached to the surface influence the band-edge position, thereby regulating the charge injection properties. From single-carrier data, it is furthermore clear that charge transport properties are also dictated by these ligands, notably in the case of native ligands, where hole transport is impeded with respect to electrons. Our DFT calculations indicate that OA is not an ideal passivating ligand for PeLED applications, due to the presence of hole traps resulting from the carboxylate group binding to Cs\textsuperscript{+}. We also demonstrated that the ligand exchange process using DDABr is reproducible, yielding minimal yet sufficient coverage for LED applications. This ligand exchange improves hole transport and enhances valence band alignment with the HOMO of the hole transport layer favorably. Together with the use of a polar electron transport layer, DDABr NC LEDs yield a more stable EQE over a broad range of current densities. These findings highlight the role of ligands and a polar ETL in modulating the properties of CsPbBr\textsubscript{3} based NCs for efficient LED applications.
\section{Experimental}

    \subsection{Sample preparation}

         \subsubsection{Materials}

            ITO (indium tin oxide) substrates with dimensions 2\,cm × 2\,cm were purchased from Kintec (Hong Kong) with a layer thickness of 90 nm on a 23 nm SiO\textsubscript{2} buffer sitting on a 0.7 mm thick glass substrate. PEDOT:PSS in a low-conductive formulation CH8000 was purchased from Heraeus Germany GmbH \& Co. KG. CsPbBr\textsubscript{3} solution (10 mg/mL in toluene, product ID 900746), DDABr (98\% purity, product ID 359025) and ZnO nanoparticle ink (2.5 wt. \%, viscosity 2.1 cP, product ID 808253) were purchased from Merck Germany KGaA. PTAA (MW = 56 kDa, PDI = 2.87) was purchased from Ossila B.V. TPBi (98\%, product ID LT-E302H) was purchased from Lumtec, Taiwan.

         \subsubsection{Ligand exchange}
         
             \begin{itemize}
                 \item Equal volumes (X) of 10 mg/ml CsPbBr\textsubscript{3} NC solution and ethyl acetate solution were mixed and centrifuged at 12 RCF for 10 minutes and the supernatant was discarded.
                 \item X volume of toluene was added to the precipitate and redispersed with X volume of ethyl acetate, centrifuged at 12 RCF for 10 minutes and the supernatant was discarded.
                 \item X volume of DDABr solution (10 mg/ml in toluene) was added along with X volume of ethyl acetate, centrifuged at 12 RCF and the supernatant was discarded. This process was repeated twice.
                \item X volume of toluene was added to the precipitate and redispersed with X volume of ethyl acetate, centrifuged at 12 RCF for 10 minutes and the supernatant was discarded.
                \item Finally X/2 volume of toluene was added to re-disperse the precipitate.
             \end{itemize}

        \subsubsection{Device fabrication}

            \label{Sample preparation}

            All samples were initially cleaned with industrial grade detergent (alconox), DI water, acetone (technical), isopropyl alcohol (technical) and isopropanol (UV grade). The samples were then treated under UV ozone for 15 minutes, immediately followed by spincoating PEDOT:PSS at 4000 rpm for 30s and annealed at 135°C for 15 minutes. The samples were transferred from the cleanroom to a nitrogen filled glove box for further preparation. PTAA solution (5 mg/ml in dichlorobenzene) was spincoated at 3000 rpm for 45s and annealed at 145°C for 15 minutes. The NC solution was spincoated by a two step process (500 rpm for 30s and 2000 rpm for 5s). Samples were then transferred to a vacuum deposition chamber with a pressure less than 10\textsuperscript{-6} mbar, without exposing to air. The samples fabricated until this step were the same for LEDs, hole-only devices and PES measurements.\\
            \textbf{LEDs} \\
                 TPBi (50 nm) was evaporated at about 0.1 nm/s, followed by LiF and aluminium at 0.03 nm/s and 0.1 nm/s, respectively.\\
             \textbf{Hole-only devices}\\
                 HATCN (10 nm) was evaporated at 0.03 nm/s followed by gold at 0.1 nm/s.\\
             \textbf{PES}\\
                The samples were prepared as mentioned before on an unpatterned ITO substrate. All procedures up until the nanocrystal deposition remain the same.\\
            \textbf{Electron-only devices}\\
                ITO substrates were cleaned and transferred to the glove box. ZnO was dropcasted onto the substrate through a PTFE filter (0.45 um), spincoated at 3000 rpm for 30s and annealed at 170°C for 15 minutes. This was followed by spincoating NC solution at 500 rpm for 30s and 2000 rpm for 5 s and evaporating LiF and aluminium at 0.03 nm/s and 0.1 nm/s, respectively.\\
         \subsubsection{LED Measurements}
             jVL curves are recorded with a Keithley 2612B source meter unit (SMU) at a sweep-rate of 0.1 V/s. A photodiode of known diameter at known distance is used for luminance detection. The electroluminescence spectrum was measured with the Phelos system from Fluxim AG (Switzerland). Combined with a Lambertian approximation, the EQE was determined.

    \subsection{UPS}
        UPS was conducted using a monochromatized helium discharge lamp with a photon energy at 21.22 eV (HIS 13 FOCUS GmbH) and a hemispherical SPECSPhoibos 100 analyzer in an ultrahigh vacuum (UHV) system (base pressure of 1 x 10\textsuperscript{-9} mbar). The monochromator eliminates the visible light and further reduces the UV flux by  a factor of 100 as compared to that of the standard helium lamp. The overall energy resolution was set at 110 meV for the UPS measurements.

    \subsection{SEC}

        SEC PL measurements were conducted in a specific home-built electrochemistry cell. As working electrode, a transparent fluorine-doped tin oxide glass substrate (20 ohm/sq, 1.1 mm thickness, Dyenamo) was used. On top of that, the NCs were drop-cast in a nitrogen-filled glovebox. The cell assembly was conducted at air. As reference electrode, an Ag wire was used and as counter electrode a Pt coil. The 0.1 M electrolyte solution was prepared in the glovebox out of five times recrystallized tetrabutylammonium hexafluorophosphate (TBAHFP, 98\% AlfaAesar), ferrocene (98\%, Acros Organics) and propylene carbonate (PC, 99.5\%, Thermo Scientific). For electrochemistry, the electrodes were connected to a BAS50 potentiostat (Bioanalytical Systems Inc.).
        For SEC, the electrochemical cell was placed on top of a confocal laser setup. The NCs were excited by a 405 nm laser (PicoQuant LDH series). The laser passed optical density filter and was focused by an air objective (Olympus LMPlan FLN, 50x, 0.5 NA) on to the sample. After excitation, the laser intensity was filtered out by a 485 nm long pass filter. Spectra were recorded by an Acton SP300i spectrometer.
        For all measurements, the open circuit potential was determined and iR compensation was applied by the potentiostat. Measurements started at approx. the open circuit potential and the integration time of the spectrometer was set to 1 s.
        

    \subsection{Simulation}
    
        All the DFT calculations have been carried out using the CP2K (version 9.1) software package\cite{Kuehne2020}. A combination of gaussian basis sets with plane waves as auxiliary basis sets known as the gaussian plane waves (GPW) method was chosen to solve the equations efficiently during all the electronic structure calculations. Double-zeta-valence-polarized (DZVP-MOLOPT-SR-GTH) basis sets\cite{VandeVondele2007} along with Goedecker-Teter-Hutter\cite{Goedecker1996} pseudopotentials for all the elements were used in combination with Perdew–Burke–Ernzerhof (PBE)\cite{Perdew1996} generalized gradient approximation (GGA) functional. Grimme’s DFT-D3\cite{Grimme2010} correction protocol was also performed to account for the van der Waals interactions. 
        Periodic boundary conditions were invoked during geometry optimization and single-point calculations, and a large cubic box size of 75\AA x75\AA x75\AA\ dimensions where the CsPbBr\textsubscript{3} nanocluster was placed inside to avoid spurious interactions between the periodic images. All the structures were optimized using the Broyden–Fletcher–Goldfarb–Shanno (BFGS) optimizer, with the maximum force criteria set to 1x10\textsuperscript{-4} Hartree/Bohr ($\sim$ 5 meV/\AA). The grid cutoff for the gaussian plane waves was chosen to be 350 Ry during all the geometry optimisation and single-point calculations, and it was increased to 450 Ry during density-of-states (DOS) calculations\cite{Lapkin2022,Wahl2022}. The electronic temperature was set to 300K and Fermi-Dirac distribution was employed during DOS calculations. All the self-consistent field (SCF) iterations were converged with a 10\textsuperscript{-6} a.u. cutoff. For modeling the perovskites and visualization of the molecular orbitals, Atomistic Simulation Environment GUI (ASE GUI) and VESTA\cite{Momma2011} programs were used.


\begin{acknowledgement}

This project was funded by Deutsche Forschungsgemeinschaft (DFG, German Research Foundation) within Priority Programme “Perovskite Semiconductors: From Fundamental Properties to Devices” (SPP 2196) under project no. 424708673 and by the KAUST CRG program “INSIGHT into Interfacial Charge Carrier Recombination Processes in Metal Halide Perovskite Solar Cells”. A. Sarkar and D. Andrienko would also like to thank the Max Planck Computing \& Data Facility (MPCDF/Raven) for providing supercomputing resources.

\end{acknowledgement}
\clearpage


\begin{suppinfo}

\renewcommand{\thefigure}{S\arabic{figure}}
\renewcommand{\thetable}{S\arabic{table}}
\renewcommand{\thesection}{S\arabic{section}}

\setcounter{figure}{0}
\setcounter{table}{0}
\setcounter{section}{0}

    \section{Quantitative NMR (qNMR)}
    
     The concentration of DDABr NCs in toluene before qNMR was determined. Following the procedure as in Dai et al\cite{Dai2024} with the difference of using ethylene carbonate standard (14 mM in DMSO-d\textsubscript{6}), A specific amount of NCs was dissolved in 0.5 ml of the ethylene carbonate/DMSO-d\textsubscript{6} solution and thereby decomposed. The amount of ligands was determined as 6000 ligands/NC for excess DDABr, 443 ligands/NC for once washed NCs and 226 for twice washed NCs, resulting in a ligand density of appx. 1 ligand/nm\textsuperscript{2} and 0.6 ligand/nm\textsuperscript{2} for once and twice washed NCs, respectively.
    
        \begin{figure}[!htb]
            \centering
                \includegraphics[width=1\textwidth]{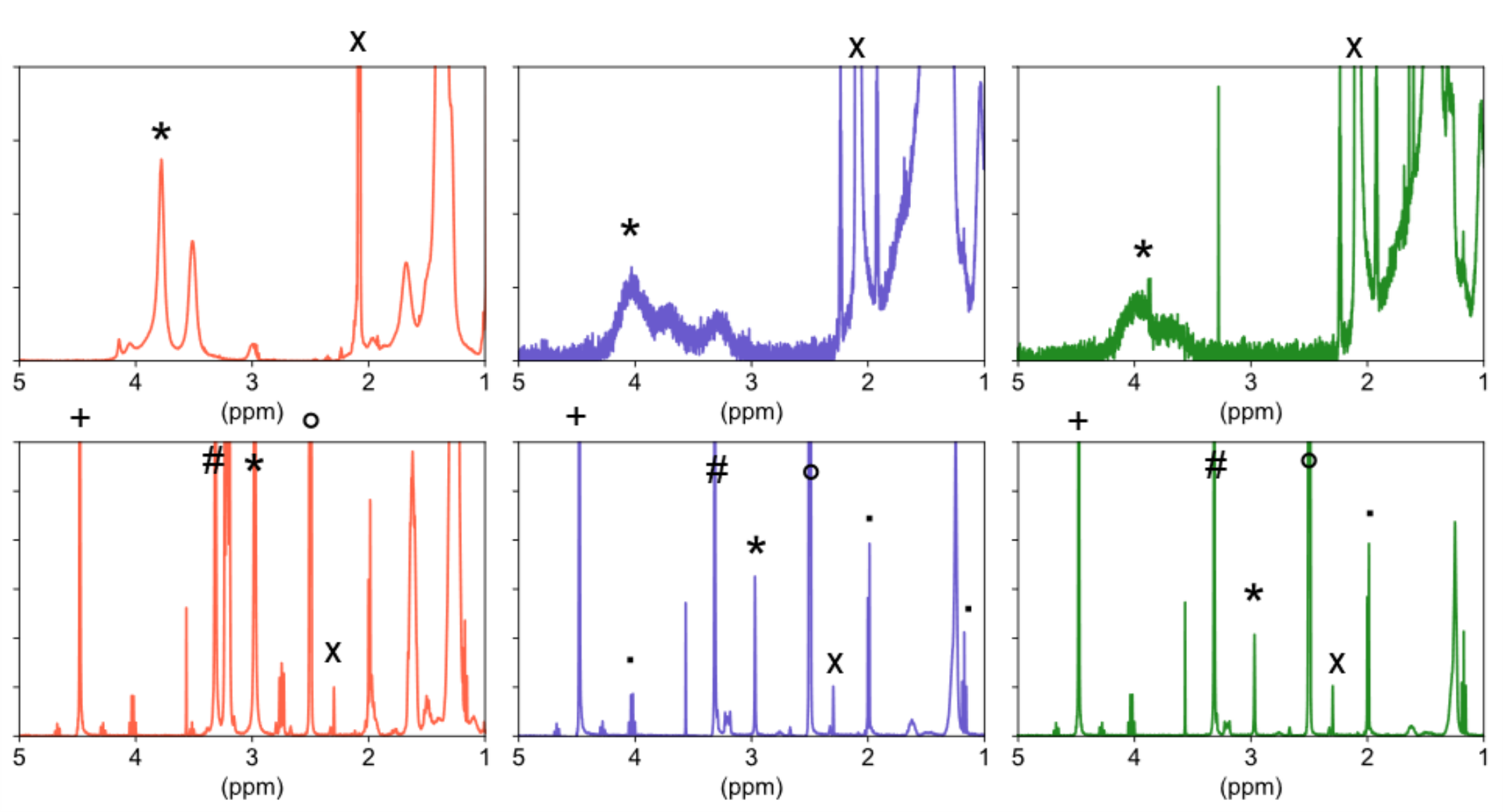}
                \caption{\textsuperscript{1}H NMR spectra of top row: bound DDABr to NCs in  toluene-d\textsubscript{8}, bottom row: quantitative NMR with ethylene glycol (+) as standard in DMSO-d\textsubscript{6}. In red: NCs with excess DDABr (methyl groups \*), in blue: NCs covered with DDABr after one washing step, in green: NCs covered with DDABr after two washing steps. Additional markers show impurities: DMSO (°, 2.5 ppm), H\textsubscript{2}O (\#, 3.33 ppm), EtOAc (., 1.99, 4.03, 1.17 ppm), toluene (x, 2.3 ppm).}
            \label{qNMR}
         \end{figure}

\clearpage

    \section{SEC}

        \subsection{Photobleach of NCs in SEC cell with electrolyte solution}

        In this section, the PL of NCs within the electrolyte solution at the open circuit potential is shown to determine the degree of photobleaching. DDABr capped NCs exhibit mostly linear and weaker photobleaching in comparison to native NCs.

            \begin{figure}[!htb]
                \centering
                    \includegraphics[width=1\textwidth]{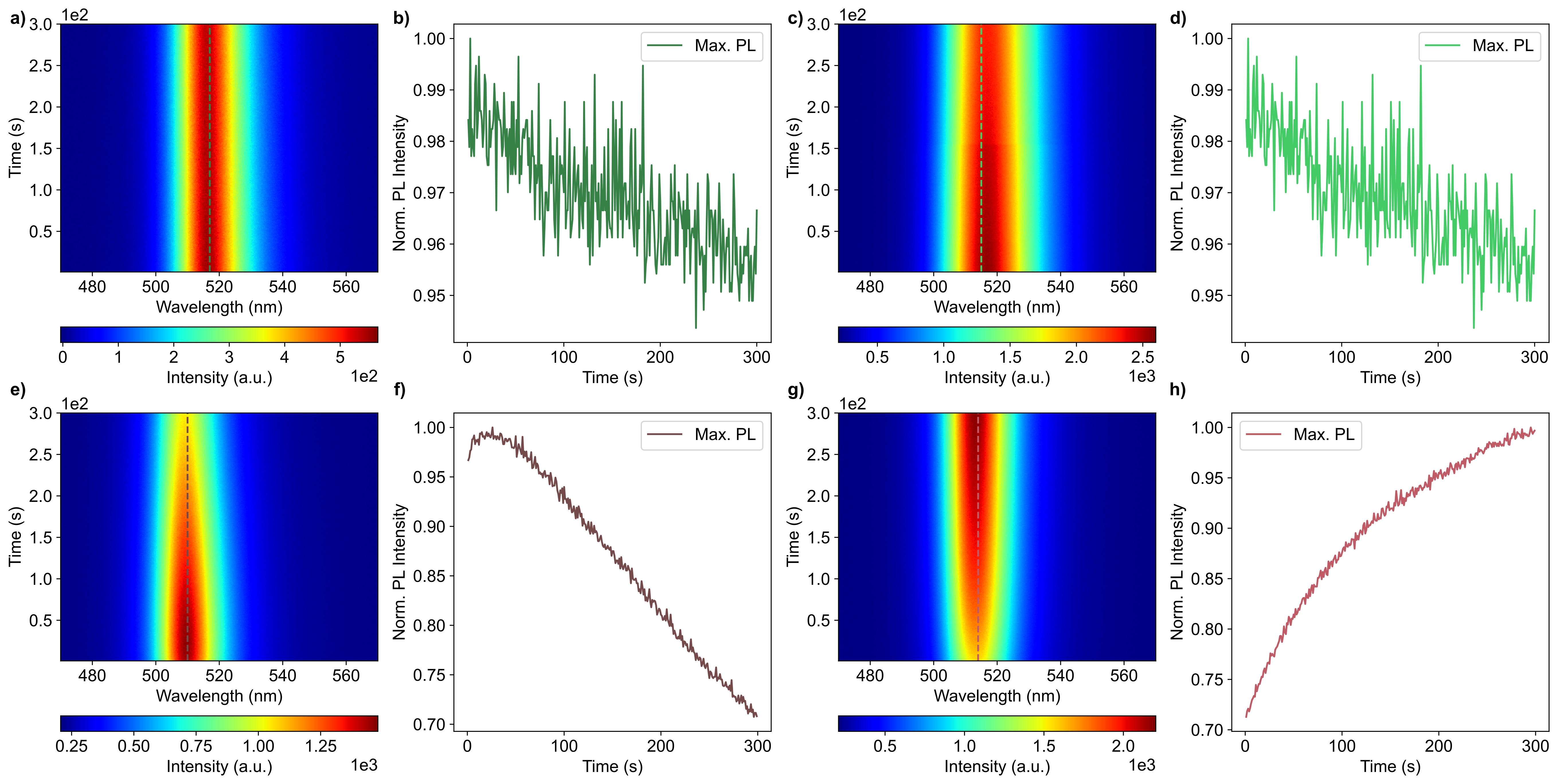}
                    \caption{PL intensity maps (on the left, a,c,e, and g) and extracted maximum PL intensity (on the right, b,d,f, and h) of DDABr and OA/OAm covered NCs with PC/TBAHFP on top for 300 s. a) and b) DDABr covered NCs after one washing step, c) and d) DDABr covered NCs after two washing steps, e) and f) OA/OAm covered NCs after one washing step, and g) and h) OA/OAm covered NCs after two washing steps.}
                \label{PL no bia}
            \end{figure}

\clearpage

        \subsection{Referencing of Ag PRE by ferrocene/ferrocenium redox couple}
        
        This section shows the CV of ferrocene in the same electrochemical cell which is used to transfer the values obtained vs. Ag PRE onto the Fc/Fc\textsuperscript{+} redox couple scale and the scale vs. vacuum in Figure 4e in the main manuscript.

            \begin{figure}[!h]
                \centering
                    \includegraphics[width=0.7\textwidth]{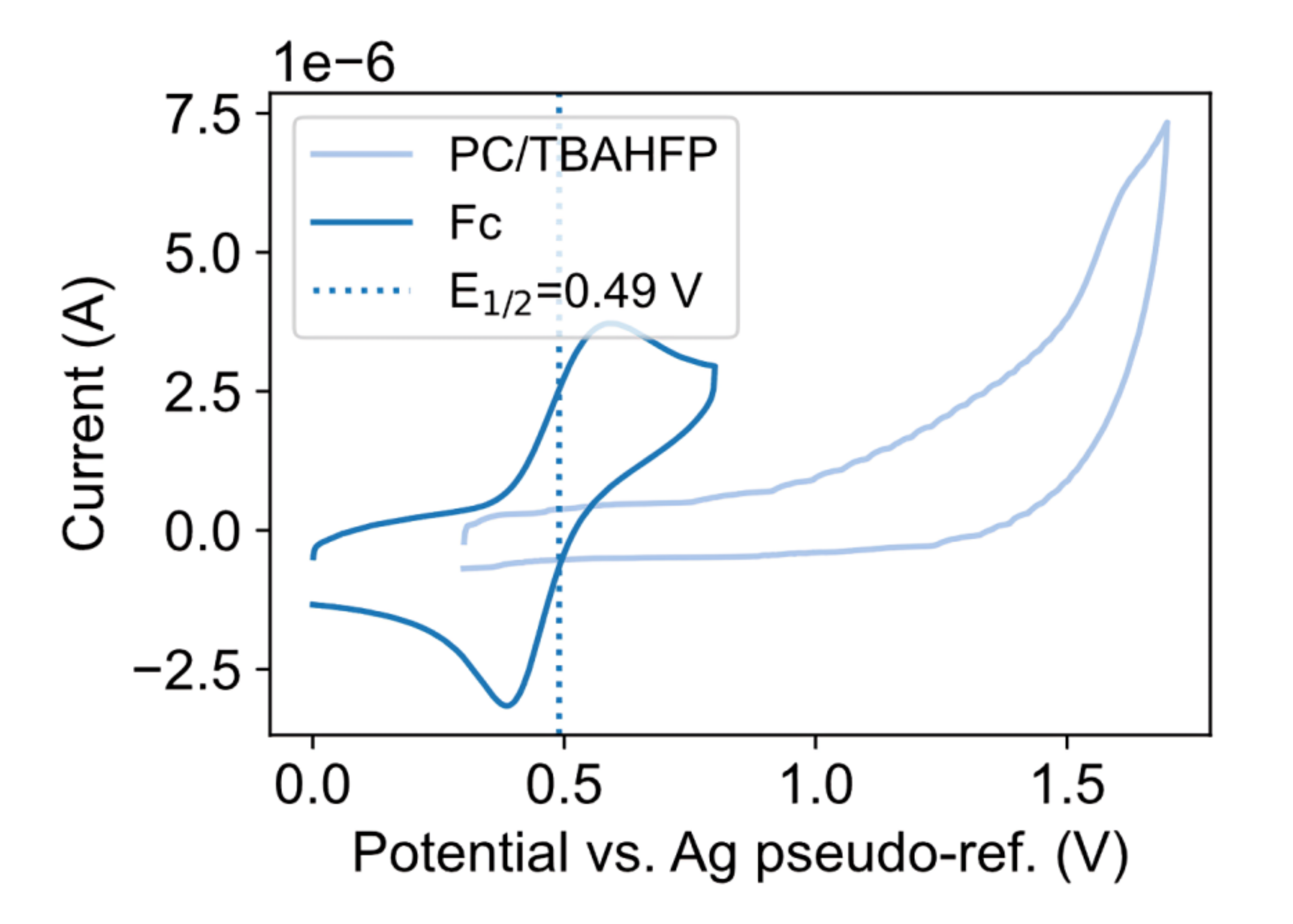}
                    \caption{Reference measurement of electrolyte solution and ferrocene/ferrocenium couple in 0.1 M PC/TBAHFP at a scan speed of 50 mV/s with half-wave potential of E\textsubscript{1/2} = 0.49 V vs. Ag PRE.}
                \label{Reference measurement}
            \end{figure}
            
        \vspace*{\fill}
        
\clearpage

        \subsection{SEC measurements of other samples in this study}
        
            \begin{figure}[!htb]
                \centering
                    \includegraphics[width=1\textwidth]{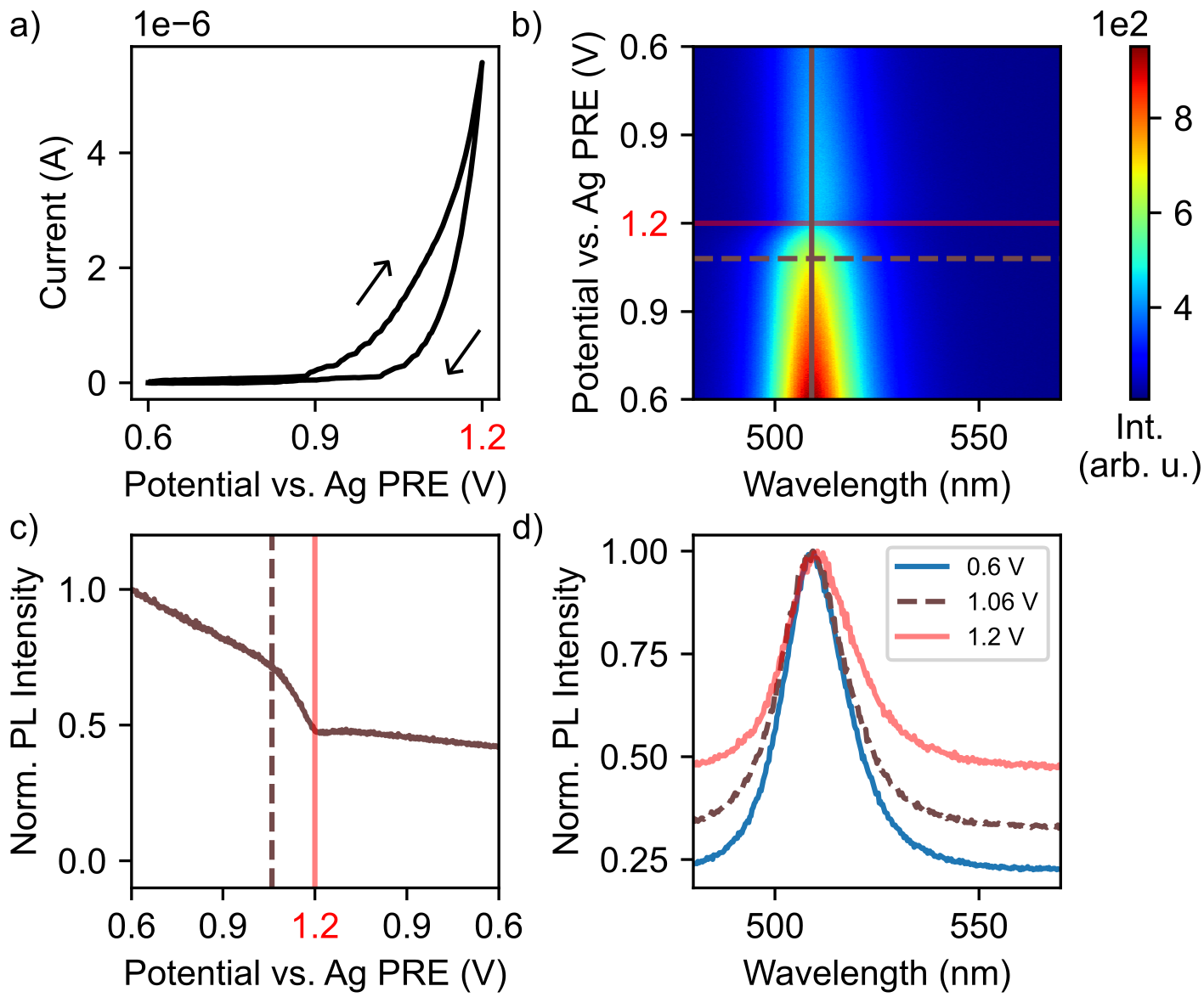}
                    \caption{SEC PL on once washed OA/OAm covered NCs in 0.1 M PC/TBAHFP with a scan speed of 2 mV/s. a) CV of NCs b) Emission spectra shown as intensity map correlated to every potential,   c) extraction of normalized PL maximum at 512 nm of the NCs, and d) single spectra extracted at the specified lines from (b)}
                \label{OA-once-precipitated}
            \end{figure}

            \begin{figure}[!htb]
                \centering
                    \includegraphics[width=1\textwidth]{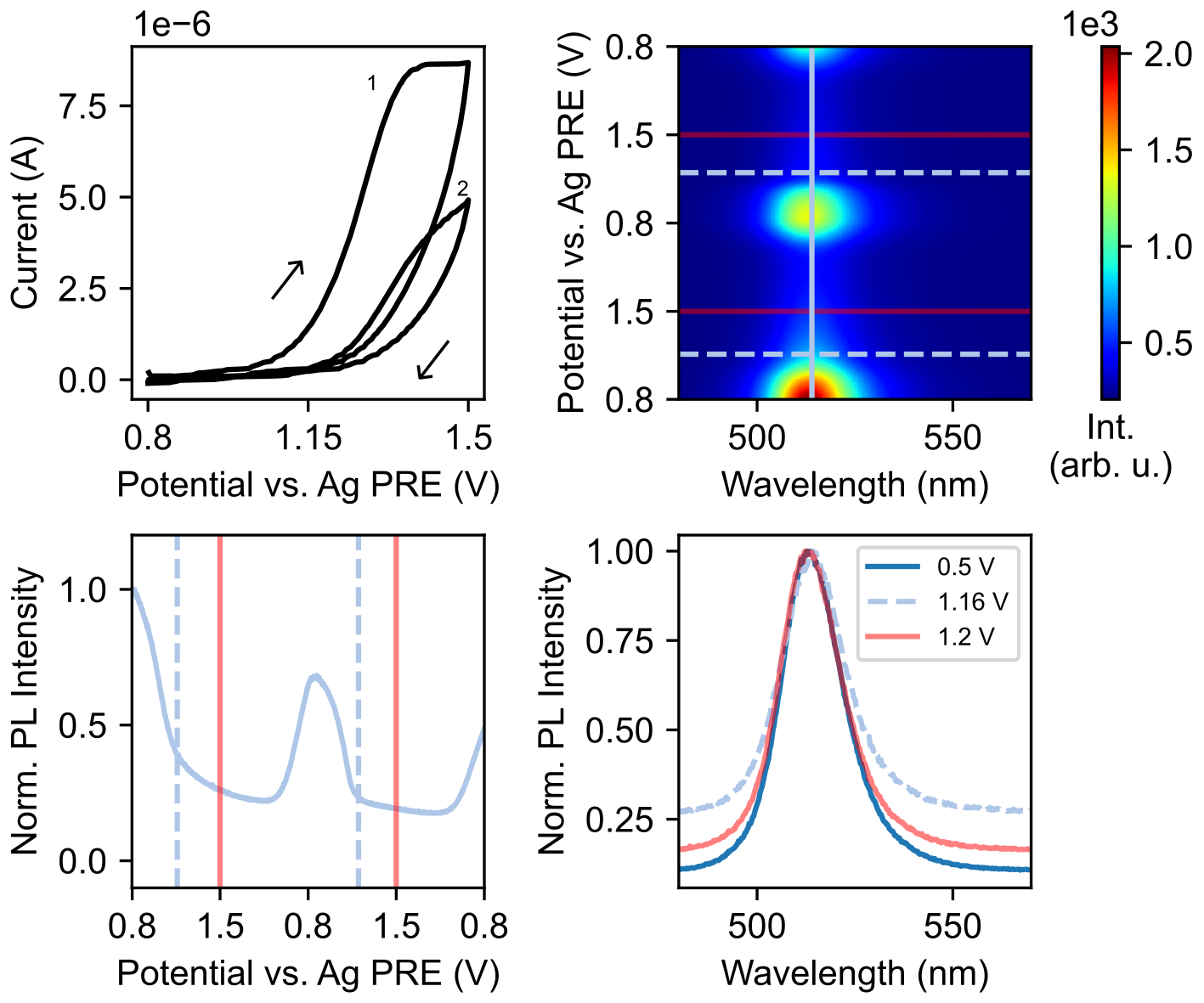}
                    \caption{SEC PL on twice washed OA/OAm covered NCs in 0.1 M PC/TBAHFP with a scan speed of 2 mV/s. a) CV of NCs,  b) Emission spectra shown as intensity map correlated to every potential, c) extraction of normalized PL maximum at 512 nm of the NCs, and d) single spectra extracted at the specified lines from (b)}
                \label{OA-twice-precipitated}
            \end{figure}

            \begin{figure}[!htb]
                \centering
                    \includegraphics[width=1\textwidth]{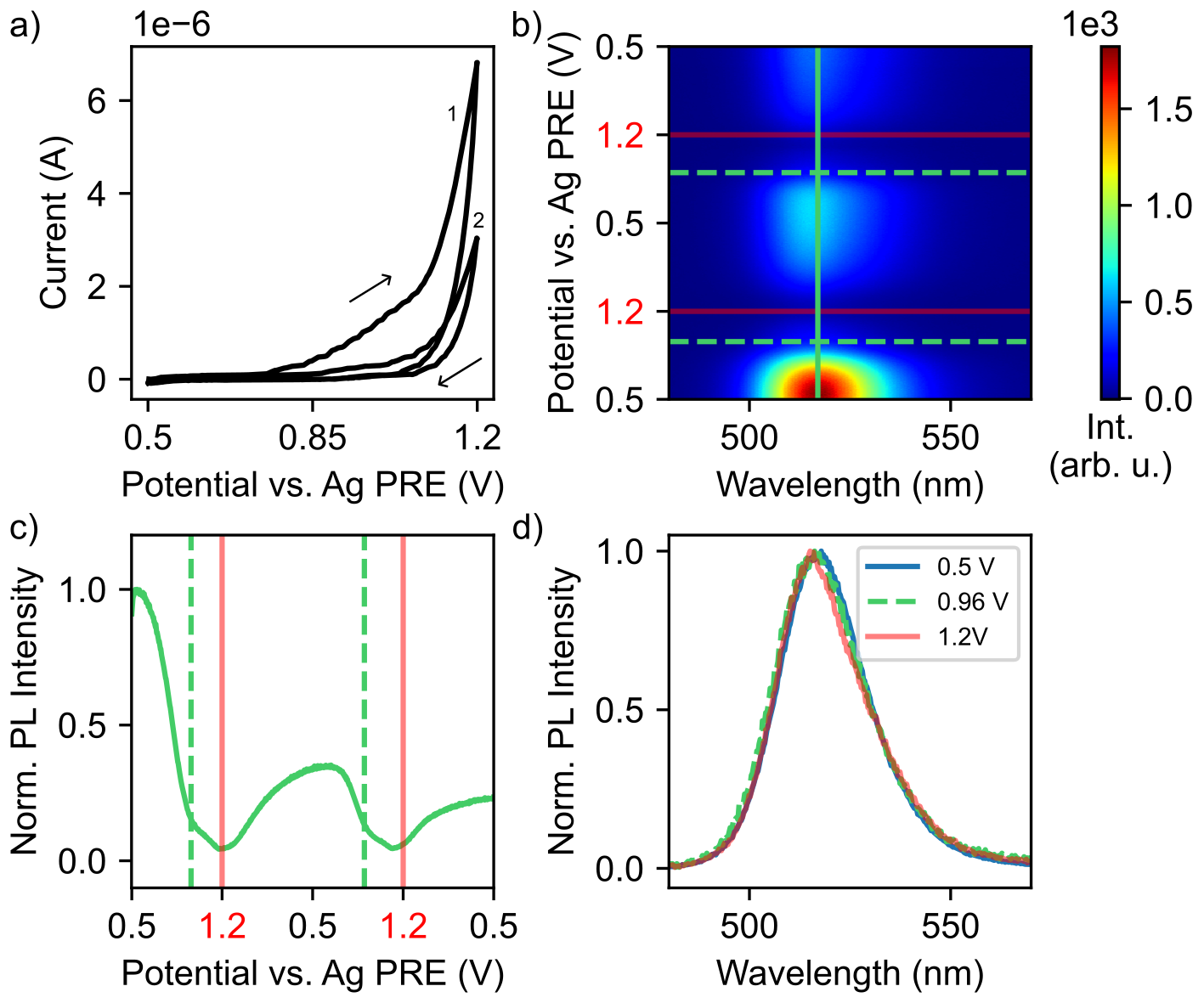}
                    \caption{SEC PL on twice washed DDABr covered NCs in 0.1 M PC/TBAHFP with a scan speed of 2 mV/s. a) CV of NCs, b) Emission spectra shown as intensity map correlated to every potential c) ,extraction of normalized PL maximum at 516 nm of the NCs, and d) single spectra extracted at the specified lines from (b)}
                \label{DDABr-twice-precipitated}
            \end{figure}
            
\clearpage

    \section{XPS}

         XPS measurements was performed on the same samples used for UPS measuements. XPS was performed at a JEOL JPS-9030 UHV system (base pressure of 1 x 10\textsuperscript{-9} mbar) using a monochromatized Al K$\alpha$ (1486.6 eV) radiation. All spectra were acquired at room temperature and normal emission. The overall energy resolution was set at 900 meV for these measurements.
         
            \begin{figure}[!htb]
                \centering
                \includegraphics[width=0.8\textwidth]{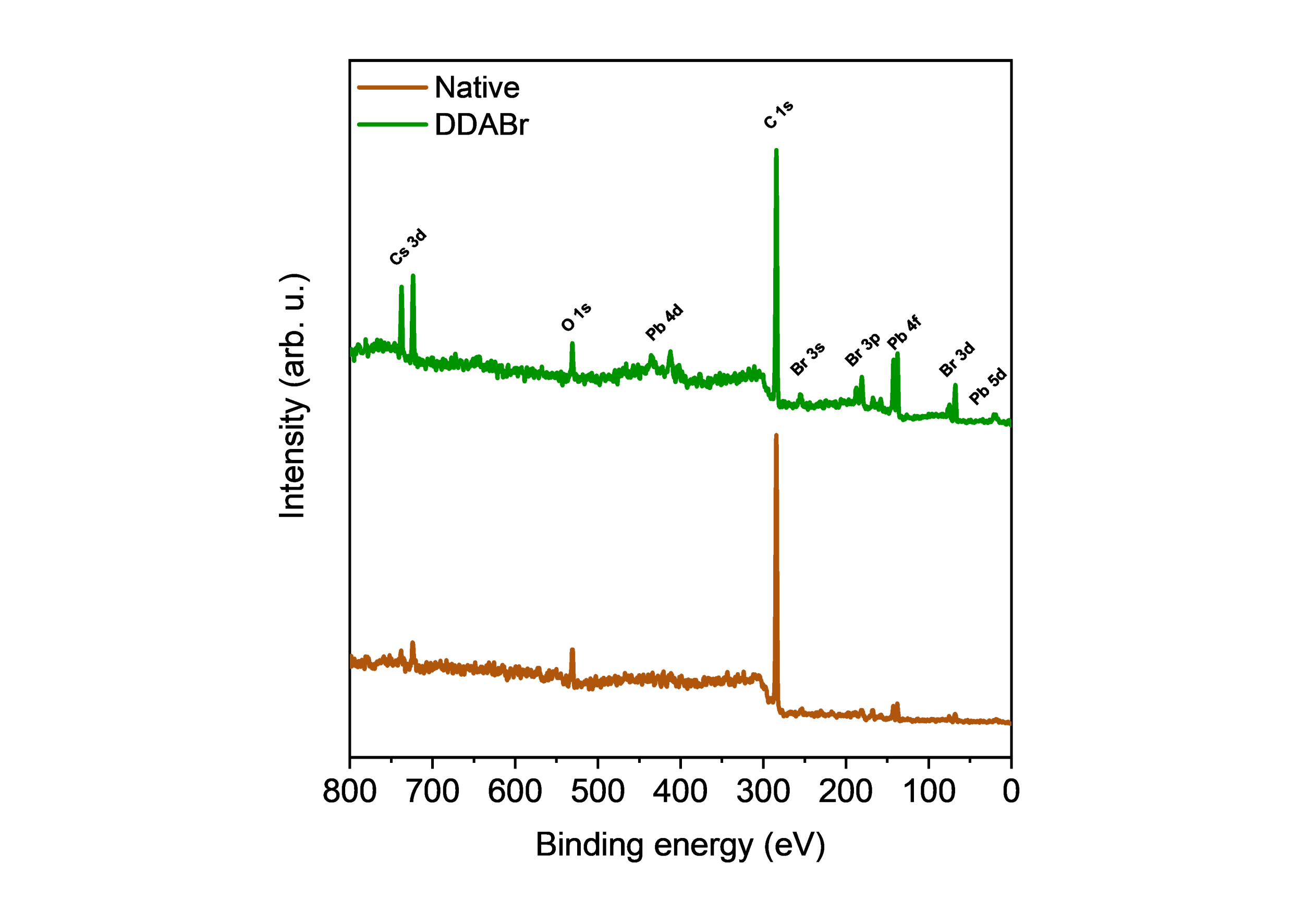}
                 \caption{XPS survey spectra of native and DDABr NCs}
                \label{XPS survey}
            \end{figure}

        \clearpage

            The elemental composition was determined by fitting the characteristic spectral peaks with Gaussian functions and quantifying the integrated peak areas.

            \begin{figure}[!htb]
                \centering
                    \subfigure[]{\includegraphics[width=0.35\textwidth]{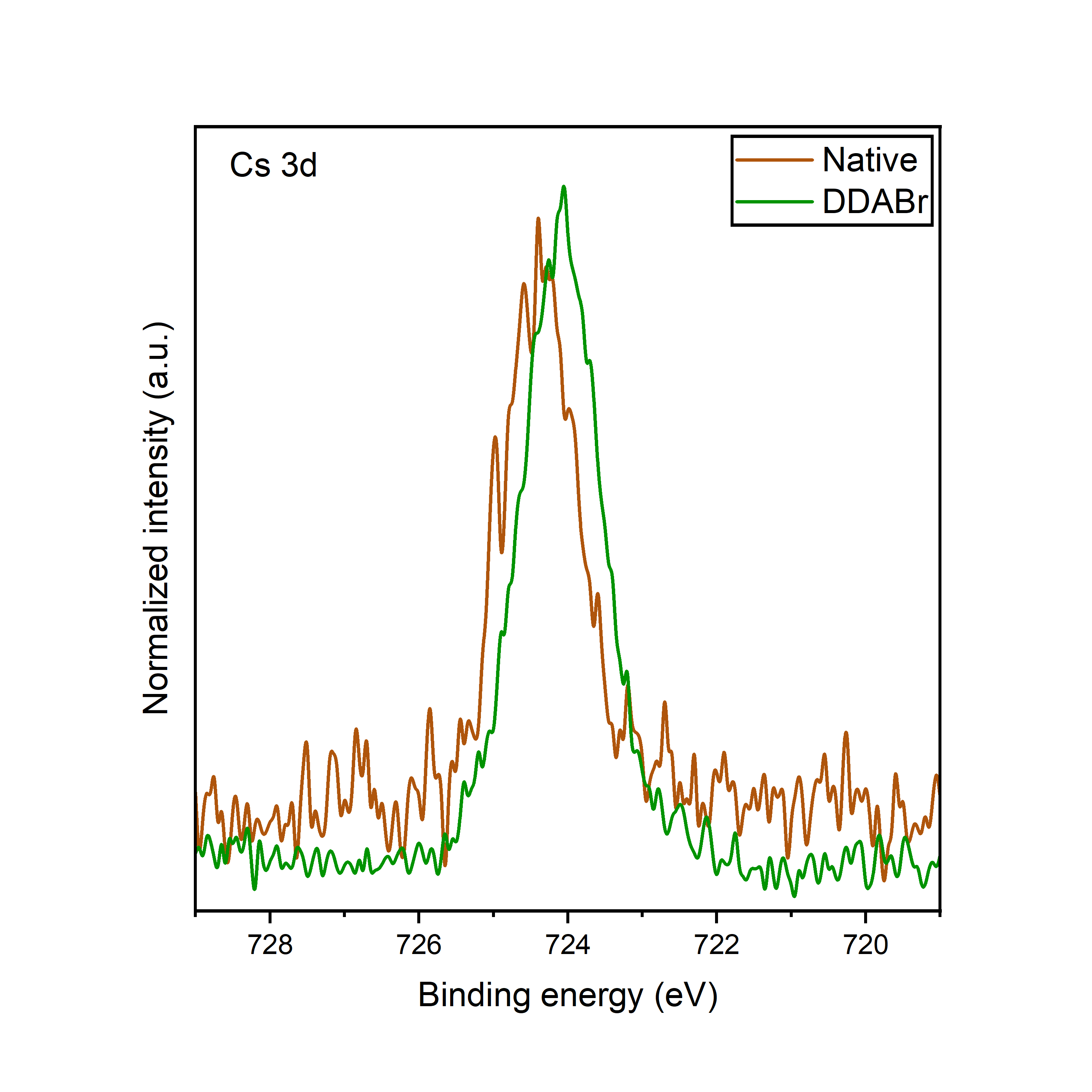}}
                    \subfigure[]{\includegraphics[width=0.35\textwidth]{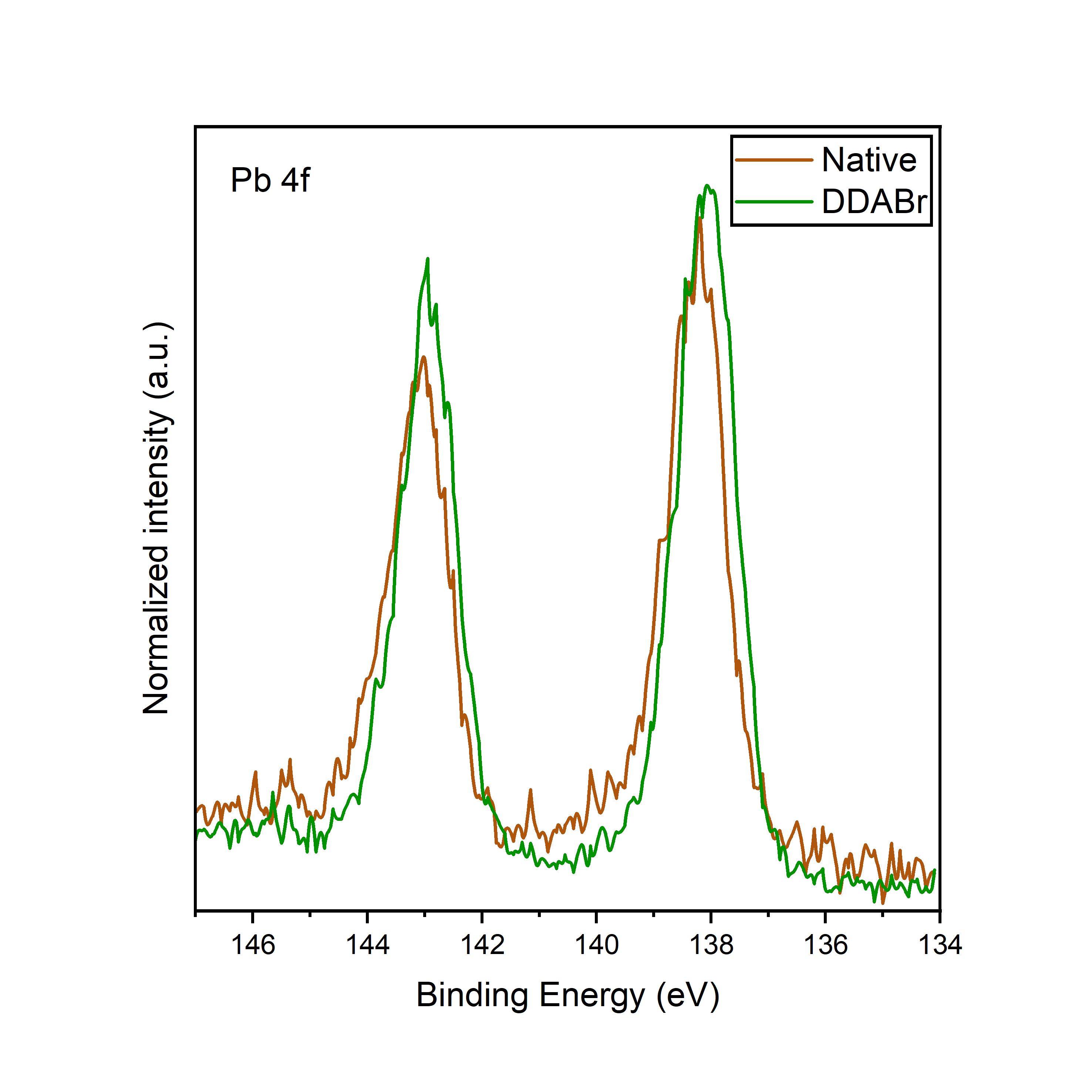} }
                    \subfigure[]{\includegraphics[width=0.35\textwidth]{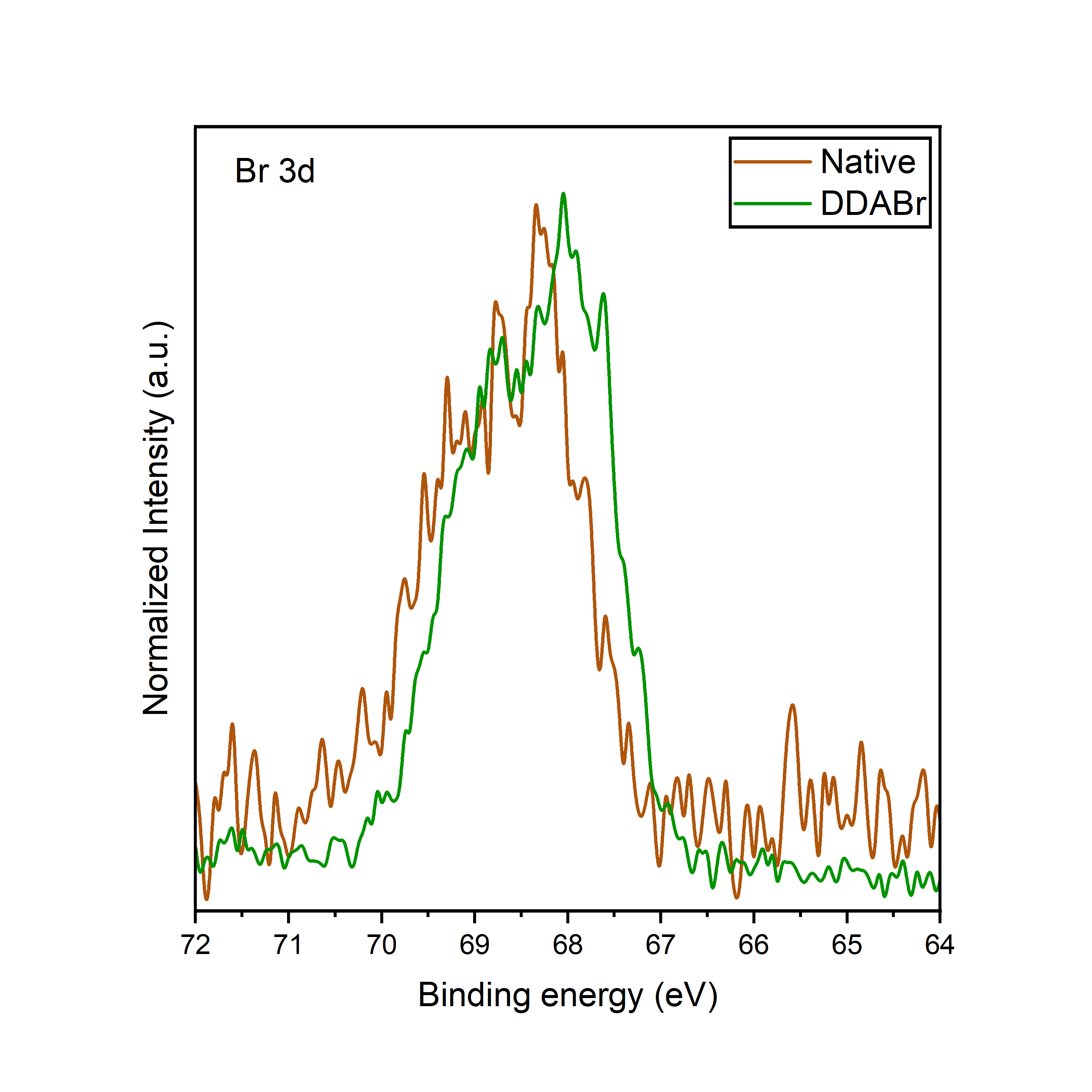} }
                \caption{XPS core levels of the individual elements in native and DDABr NCs}
                \label{XPS elements}
            \end{figure}

            \begin{table}[!htb]
                \centering
                     \begin{tabular}{cc|c|c|c|c|l}
                        \\ \cline{3-6}
                        & & Pb 4f 7/2  &  Br 3d 5/2 & Cs 3d 5/2 &  Br/Pb \\
                        \cline{1-6}
                        \multicolumn{1}{ |c  }{\multirow{2}{*}{Relative atomic ratio} } &
                        \multicolumn{1}{ |c| }{Native} & 27.1 \% & 50.2 \% & 22.7 \% & 1.85 &     \\ 
                        \cline{2-6}
                        \multicolumn{1}{ |c  }{}                        &
                        \multicolumn{1}{ |c| }{DDABr} & 21.7 \% & 55.8 \% & 22.5 \% & 2.57 &     \\
                        \cline{1-6}
                    \end{tabular}
                \caption{Relative atomic ratios calculated by integrating XPS core level spectra for native and DDABr NCs.}
                \label{RAR}
            \end{table}

\clearpage

    \section{Additional DFT calculations}

            \begin{figure}[!htb]
                \centering
                    \includegraphics[width=1\textwidth]{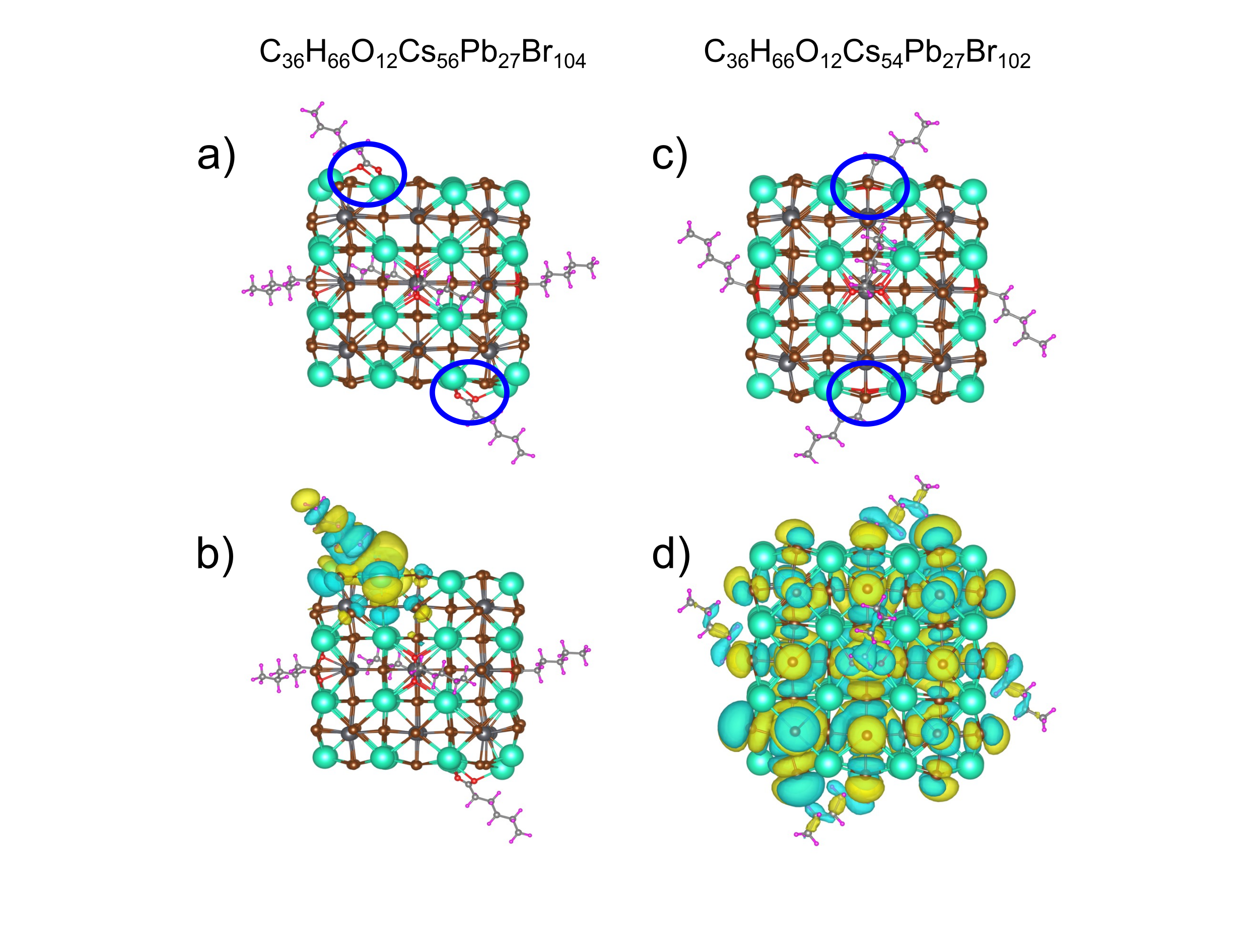}
                    \caption{Two different binding modes of the carboxylate oxygen atoms in the hexanoate (mimicking the oleate ligand) group which are attached to the PNC surface (encircled). a and b are the cases where two of the carboxylate oxygens are bonded to the Cs\textsuperscript{+} ions. c and d represents the other less likely binding modes where all the carboxylate oxygens are directly bonded to the Pb\textsuperscript{2+} sites.}
                \label{One oleate HOMO}
            \end{figure}
\clearpage

    \section{Kelvin probe measurements}

            \begin{figure}[!htb]
                \centering
                    \includegraphics[width=1\textwidth]{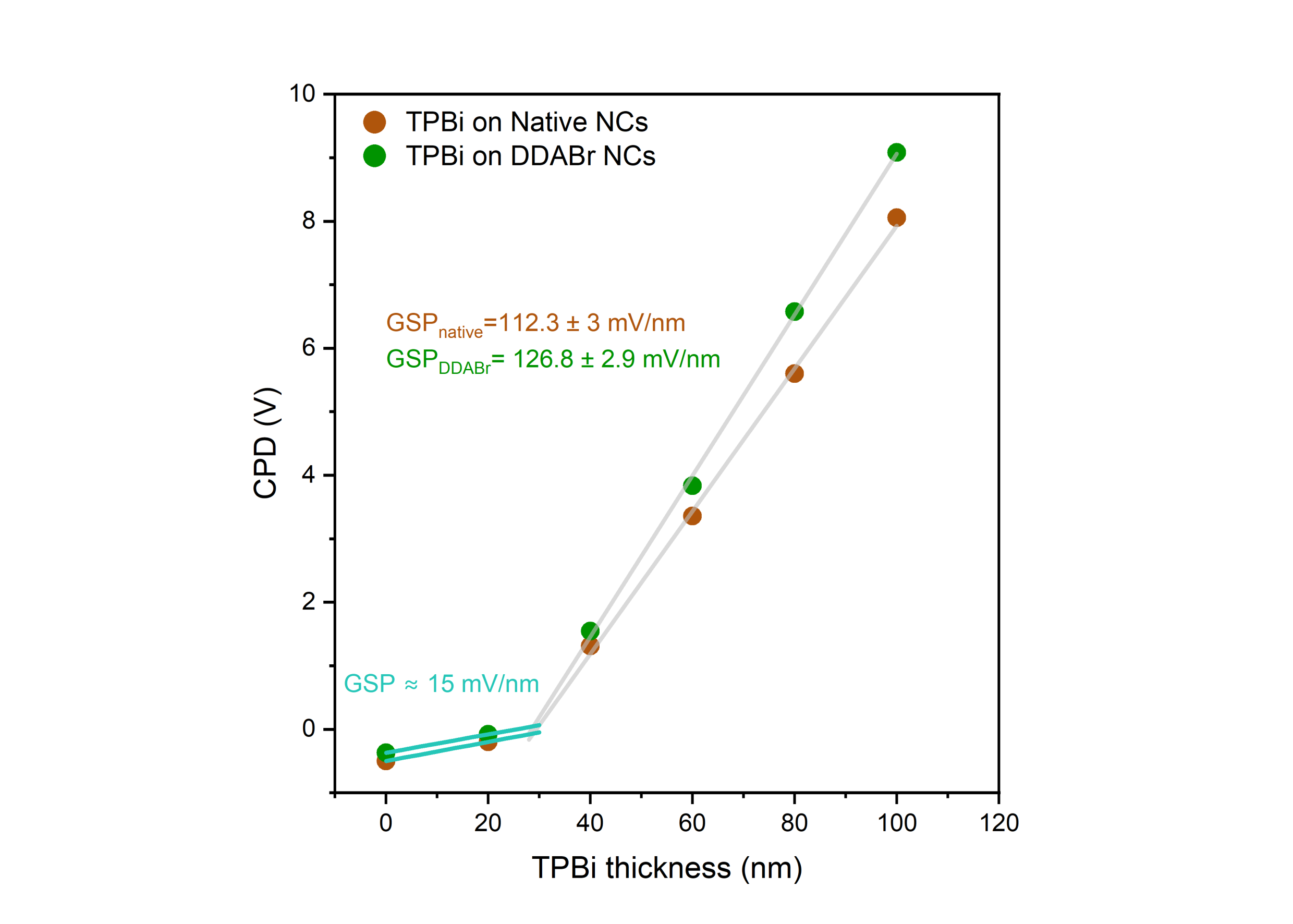}
                    \caption{Contact potential difference (CPD) measured from Kelvin probe vs thickness of TPBi evaporated on native (brown circles) and DDABr (green circles). GSP is extracted from the slope of the resulting linear fit on plotted data.}
                \label{KP}
            \end{figure}

\clearpage

    \section{Washed vs unwashed native LEDs}

            \begin{figure}[!htb]
                \centering
                    \includegraphics[width=1\textwidth]{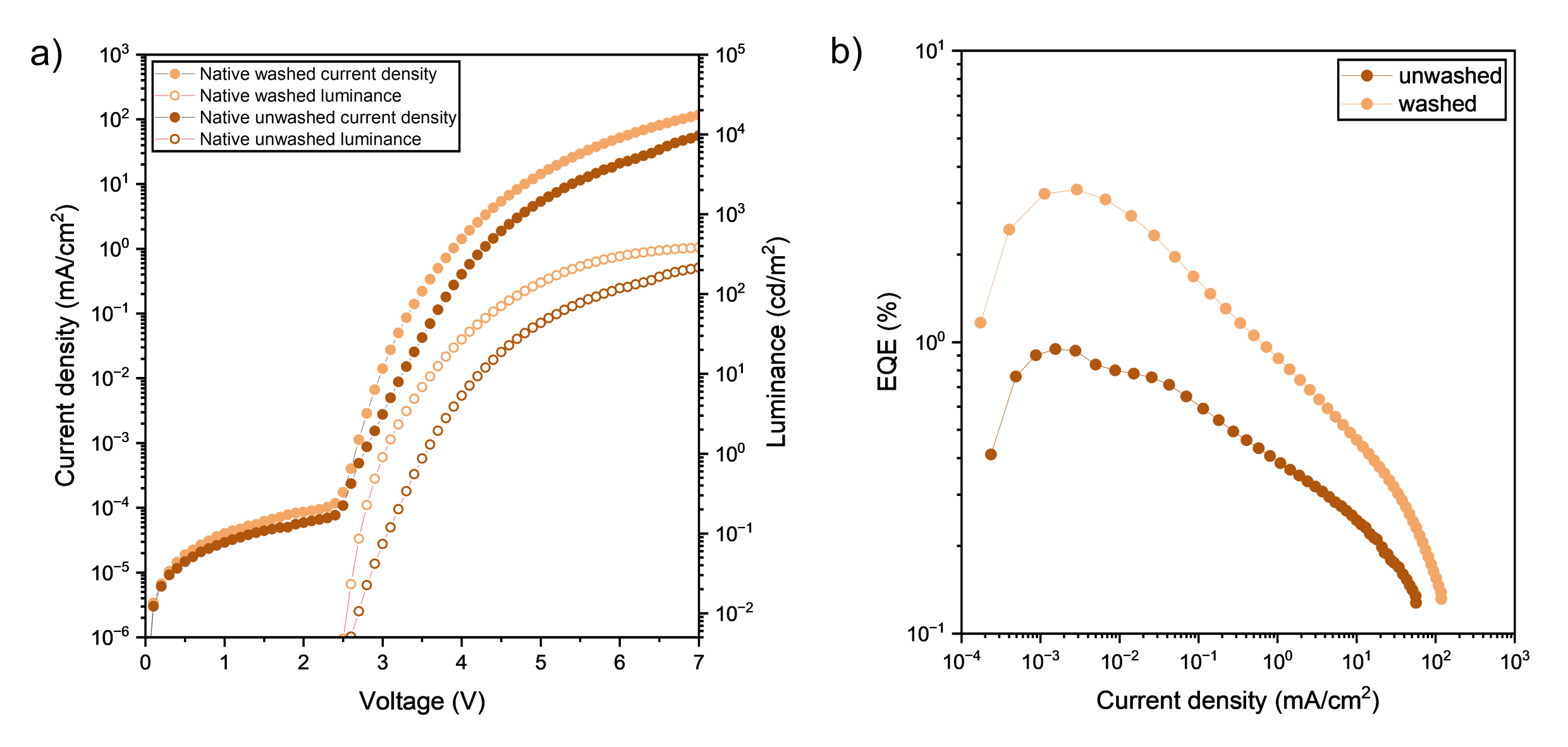}
                    \caption{J-V-L and EQE plots of Washed vs unwashed native NC LEDs}
                \label{Washed vs unwashed}
            \end{figure}

\clearpage

     \clearpage

\end{suppinfo}


\bibliography{References}

\end{document}